\documentclass[final,authoryear,5p,twocolumn]{elsarticle}

\journal{Physics of the Dark Universe}

\usepackage[T1]{fontenc}
\usepackage[usenames]{xcolor}
\usepackage[colorlinks=true]{hyperref}
\usepackage{amssymb}

\usepackage{amsmath}	    
\usepackage{tikz}           
\usetikzlibrary{positioning}

\graphicspath{{./figs/}}

\newcommand{\natastron}{Nat.Aston.}
\newcommand\aap{A\&A}                
\newcommand\aj{AJ}                   
\newcommand\apj{ApJ}                 
\newcommand\apjl{ApJ}                
\newcommand\mnras{MNRAS}             

\newcommand{\mscript}[1]{{\mbox{\scriptsize \rm #1}}}
\newcommand{\rn}{{r_\mscript{n}}}
\newcommand{\logYD}{{\Upsilon_\mscript{d}}}
\newcommand{\logYB}{{\Upsilon_\mscript{b}}}
\newcommand{\YD}{{Y_\mscript{d}}}
\newcommand{\YB}{{Y_\mscript{b}}}
\newcommand{\VD}{{V^2_\mscript{*d}}}
\newcommand{\VB}{{V^2_\mscript{*b}}}

\newcommand{\Vgas}{{V^2_\mscript{gas}}}
\newcommand{\Vbar}{V^2_{\mscript{bar}}}
\newcommand{\Vmod}{V^2_{\mscript{mod}}}
\newcommand{\Vobs}{V^2_{\mscript{obs}}}
\newcommand{\lambdaO}[1]{{\lambda_\mscript{obs}^\mscript{(#1)}}}

\begin{document}

\begin{frontmatter}

\title{Normalized additional velocity distribution: testing the radial profile of dark matter halos and MOND}

\author[dfis,center,heidelberg]{Davi C. Rodrigues}
\ead{davi.rodrigues@ufes.br}
\author[center,heidelberg]{Alejandro Hernandez-Arboleda}
\ead{alejandro.arboleda@aluno.ufes.br}
\author[aneta]{Aneta Wojnar}
\ead{aneta.magdalena.wojnar@ut.ee}

\affiliation[dfis]{
  organization={Departamento de Física, Universidade Federal do Esp\'irito Santo},
  city={Vit\'oria},
  state={ES},
  country={Brazil}
}

\affiliation[center]{
  organization={N{\'u}cleo de Astrof\'isica e Cosmologia \& PPGCosmo,  Universidade Federal do Esp{\'i}rito Santo},
  city={Vit\'oria},
  state={ES},
  country={Brazil}
}

\affiliation[heidelberg]{
  organization={Institute for Theoretical Physics, University of Heidelberg},
  addressline={Philosophenweg 16, D-69120}, 
  city={Heidelberg},
  country={Germany}
}

\affiliation[aneta]{
  organization={Departamento de F\'isica Te\'orica, Universidad Complutense de Madrid},
  addressline={E-28040}, 
  city={Madrid},
  country={Spain}
}

\begin{abstract}
We propose a complementary and fast approach to study galaxy rotation curves directly from the sample data, instead of individual fits.
With this approach, some relevant tests  can be done analytically.
It is based on a dimensionless difference between the observational rotation curve and the expected one from the baryonic matter ($\delta V^2$) as a function of the normalized radius $\rn$ (i.e., for all galaxies, $0 < \rn < 1$). 
Using 153 galaxies from the SPARC galaxy sample, we find the observational distribution of $\delta V^2$. 
Considering radii with $0.2 < \rn < 0.9$, most of the SPARC data are close to the curve $\delta V^2 = \rn^{0.42}$, and about $95\%$ of the SPARC data is between the curves $\delta V^2 = \rn^{2.2}$ and $\delta V^2 = 2 \rn^{0.38} - \rn^{1.9} $. 
We consider three well known dark matter halo models (NFW, Burkert and DC14), a simple dark matter rotation curve profile for the purpose of model comparison (Arctan$_\alpha$) and one modified gravity model without dark matter (MOND). 
By comparing the observational data distribution with the model-inferred data, we confirm that the NFW halo lacks the necessary diversity to reproduce several observed rotation curves, while Burkert and DC14 models have better concordance with observational data.
The lowest $\delta V^2$ curves that can be found from NFW are linear on the normalized radius (i.e., $\delta V^2_\mscript{NFW} = \rn$), while for Burkert $\delta V^2_\mscript{Bur} = \rn^2$ (this result is independent of the halo density parameter, i.e., $\rho_{\rm c}$ or $\rho_{\rm s}$). 
MOND only covers the very central region of the observed distribution, hence it also lacks the necessary diversity, which in turn is related to larger $\chi^2$ values. 
In a second paper, the method will  be extended to consider other classes of modified gravity models.

\end{abstract}


\end{frontmatter}


\section{Introduction}

Galaxies constitute one important and unique piece of information about dark matter-related phenomena. The standard approach, considering the agreement between several independent probes \citep{Bergstrom:2000pn, Courteau:2013cjm, Bertone:2016nfn}, is to assume dark matter existence at cosmological scales and that the cosmological evolution leads to the dark matter distribution found in galaxies. At galactic scales, a relevant complication in this picture is the baryonic feedback, which in general has a relevant impact on the dark matter distribution. 

There are currently different dark matter halo profiles and also some modified gravity models that are candidates to explain dark matter-like phenomena in galaxies. Here we consider four well-known possibilities, namely: $i$) the Navarro-Frenk-While (NFW) \citep{Navarro:1996gj} profile, which is inferred from N-body dark matter simulations (thus without baryonic feedback); $ii$) the phenomenological Burkert profile \citep{1995ApJ...447L..25B}, which has a core and it has good performance on fitting several galaxies \citep[e.g.,][]{Salucci:2000ps, deBlok:2001fe, Rodrigues:2017vto, Li:2020iib}; $iii$) the \citet{DiCintio:2014xia} halo (DC14), which considers hydrodynamical simulations with baryonic feedback \citep[see also][]{Katz:2016hyb}; and $iv$) Modified Newtonian Dynamics (MOND) \citep{1983ApJ...270..371M, Famaey:2011kh} which is a well-known and simple modified gravity model that is capable of reproducing relevant trends observed in galaxy rotation curves \cite[e.g.,][]{McGaugh:2016leg}. For comparison purposes, we also consider a less cited phenomenological possibility based on an $\arctan$ function, here named Arctan$_\alpha$, which will be introduced latter.

In this work we develop and apply a method that tests a given model with a sample of galaxies, instead of doing galaxy by galaxy fits. The method aims to test both dark matter and modified gravity models. Instead of trying to test all the properties simultaneously, we focus on the radial dependence of the dark matter contribution to the rotation curve, or the modified gravity one. 

First we introduce the squared additional velocity, denoted by $\Delta V^2$, which is the contribution to the circular velocity that is not due to the expected Newtonian baryonic contribution.\footnote{In the context of Newtonian gravity with dark matter, this quantity was named the rotation velocity attributed to dark matter ($V_h$) \citep{McGaugh:2006vv}.} We use the squared velocity since in this form it can be directly added to other Newtonian contributions.\footnote{For instance, for a galaxy with stars, atomic gas and dark matter, whose Newtonian potentials are respectively denoted by $\Phi_\mscript{*}$, $\Phi_\mscript{gas}$ and $\Phi_\mscript{dm}$, the total Newtonian potential is $\Phi = \Phi_\mscript{*} + \Phi_\mscript{gas} + \Phi_\mscript{dm}$. Hence, the total circular velocity is $V^2 = V^2_\mscript{*} + V^2_\mscript{gas} + V^2_\mscript{dm}$. For this example, which is  based on Newtonian gravity, $\Delta V^2 = V^2_\mscript{dm}$.} 

In order to study the radial dependence of $\Delta V^2$,  we introduce the normalized additional velocity (NAV), denoted by $\delta V^2$. The NAV is the core quantity here studied. It commonly reduces the number of model's free parameters by one and allows for additional analytical estimates before doing numerical rotation curves (RCs) evaluations. Moreover, instead of studying each galaxy individually  to understand if a model is viable or not, it is possible to compare the observational with the model NAV distributions, as we show here.

This work is organized as follows: in the next section we present the method, explaining how to compute $\delta V^2(\rn)$ from the observational data and from a model. Section \ref{sec:models} is devoted to presenting the considered models and to apply the proposed method. Section \ref{sec:modelComparisons}  summarizes the model results and compare them. In Sec.~\ref{sec:conclusions}, we review and discuss our results.  \ref{app:KDE} is devoted to a brief review on KDEs and the Silverman bandwidth. In \ref{app:IndividualFits} we present our results for the the individual fits of galaxies, considering six different models. Additional details on the computations can be found in the \texttt{NAVanalysis} code \citep{NAVanalysis}. \ref{app: M200-c} comments on the NFW halo results in the presence of mass-concentration constraints ($M_{200}-c$).

\section{The method} \label{sec:method}

\subsection{Outline} \label{sec:methodOutline}

The method deals with dark matter or modified gravity models. The dark matter case: for a given dark matter model and a data sample of galaxy rotation curves, the method compares the difference between the observed rotation curve (RC) and the baryonic contribution with the dark matter rotation curve contribution. However, instead of doing this galaxy by galaxy, where the individual uncertainties are more relevant, one considers the full set of data points of the complete sample of galaxies. Moreover, to simplify the analysis and to focus on a relevant aspect of dark matter halos, this comparison is done under a normalization. For the modified gravity case: the method essence is the same, but the ``baryonic contribution'' is understood as the Newtonian baryonic contribution alone, while the non-Newtonian contribution is in place of the dark matter contribution.

The observational additional (squared) velocity, $\Delta V^2_\mscript{obs}$, is defined as
\begin{equation}
	\Delta \Vobs \equiv \Vobs - \Vbar  \, . \label{DVobs}
\end{equation}
and the model additional (squared) velocity, $\Delta V^2_\mscript{mod}$,
\begin{equation}
	\Delta \Vmod \equiv \Vmod - \Vbar \, .
\end{equation}
In the above, $V_\mscript{obs}$ is the the observational RC, $V_\mscript{mod}$ is the   model RC and $V_\mscript{bar}$ is  the expected baryonic RC (which is inferred from the stars and the HI regions radiation). To be more explicit, 
\begin{equation}
	\Vbar =  \VD + \VB + \Vgas \, ,
\end{equation}
where $\VD$, $\VB$ and $\Vgas$ are the squared circular velocity of the following components respectively: stellar disk, stellar bulge and the gas (atomic hydrogen and helium). The mass-to-light ratios are included in these quantities.

These squared velocities are not truly the square of a given physical velocity (this is a common convention), namely 
\begin{equation}
	V^2_\mscript{x}(r) \equiv	a_\mscript{x}(r) \, r \, ,
\end{equation}
that is, the component x contribution is the centripetal acceleration due to the component x times the (cylindrical) radial position $r$. Hence, since the matter distribution is not spherical, $V^2_\mscript{x}$ can be either positive or negative. A negative $V^2_\mscript{x}$ means that the component $x$, in a given radius, reduces the total centripetal acceleration. It is common to use this convention. This detail is relevant depending on the particular model and the particular galaxy considered.

\begin{figure}
	\begin{tikzpicture}
  		\node (img1)  {\includegraphics[width=0.45\textwidth]{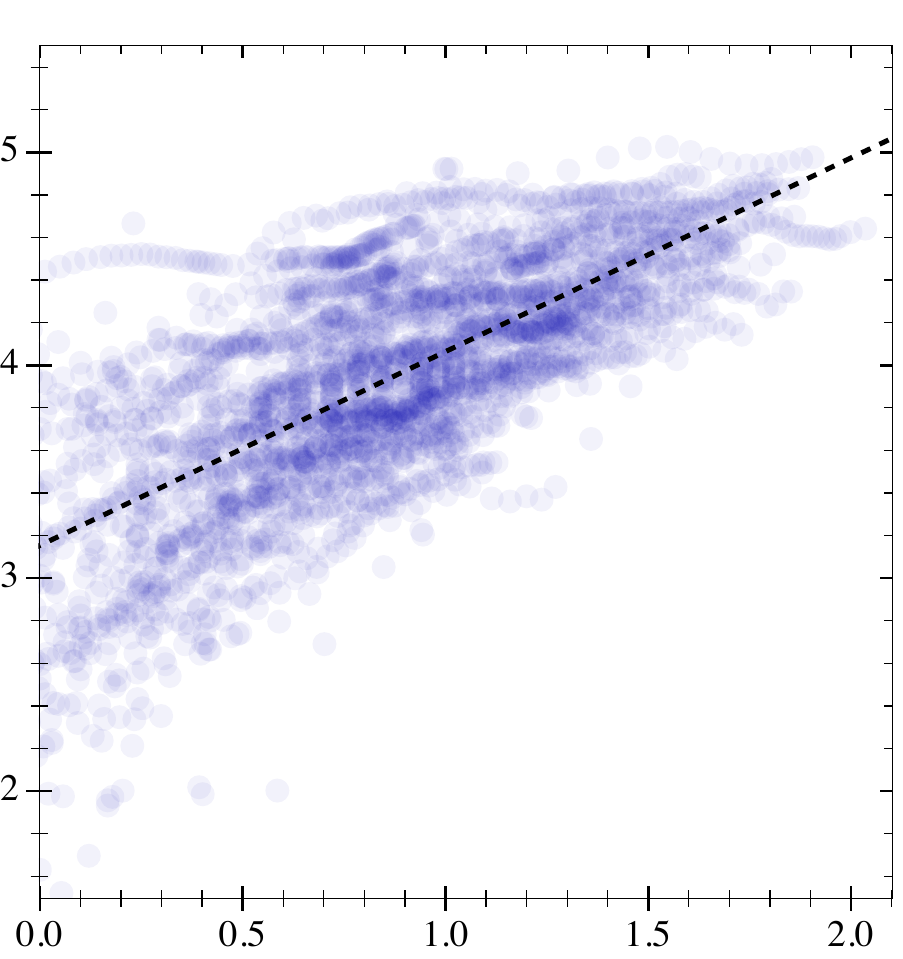}};
		\node[below=of img1, node distance=0cm, yshift=1.1cm, xshift=0.3cm, font=\color{black}] {\large $\log_{10} R \;\; ({\rm kpc})$};
		\node[left=of img1, node distance=0cm, rotate=90, yshift=-0.7cm, xshift=1.5cm] {\large $\log_{10}(\Delta \Vobs) \;\; ({\rm km}^2/{\rm s}^2)$};
	\end{tikzpicture} 
	\caption{The additional velocity ($\Delta \Vobs$) is here plotted for 153 galaxies from SPARC, and with $\YD = 0.5$ and $\YB = 0.6$. Each rotation curve data point is represented with one transparent blue disk. The dashed line is best fit result \eqref{DeltaVobsBest}.}
	\label{fig:DeltaV2Analysis}
\end{figure}

An analysis on $\Delta \Vobs$ was performed by \cite{McGaugh:2006vv}, where it was  found the following  approximate description\footnote{This is $V_h^2$, in their notation.} $\Delta \Vobs \approx  10^{2.94} \, r \; {{\rm km}^2}/{{\rm s}^2}$, for $1 < r  \, ({\rm kpc}) < 74$. This profile was found from 60 galaxies and under assumptions on the stellar mass-to-light ratios that are different from those adopted in the SPARC data. Nonetheless, for the 153 SPARC galaxies we find results that are very similar, namely 
\begin{align}
  & \Delta \Vobs \approx 10^{3.08} \, r \; {\rm km}^2/{\rm s}^2  \,  \mbox{ or,} \label{DeltaVobsBestModel0}\\[.1cm]
  & \Delta \Vobs \approx 10^{3.15}\, r^{0.91} \; {\rm km}^2 / {\rm s}^2 \, .\label{DeltaVobsBest}
\end{align}
Data with $r < 1$ kpc are neglected, since it is a region with larger dispersion. The first case considers a simpler model, in which $\Delta \Vobs$ is taken as linear on $r$, like the one adopted by \citet{McGaugh:2006vv}, while the latter \eqref{DeltaVobsBest} considers a free parameter for the $r$ power. In the end, all the results are similar. 

Figure~\ref{fig:DeltaV2Analysis} shows the considered data and plots the last curve. This figure  suggests that $\Delta \Vobs$ commonly increases with the radius. However,  we remark three issues about this plot: i) different radial regions can have large differences on the number of data points, ii) the magnitude of $\Delta \Vobs$  from two different galaxies may be very different, even though their radial dependence may be similar, and iii) if the magnitude of $\Delta \Vobs$ correlates with the radial range probed by the galaxies, the trend seen in Fig.~\ref{fig:DeltaV2Analysis} will not match, even approximately, the trend in the individual galaxies. Hence, it is not possible to infer that good models should satisfy $\Delta \Vmod \propto r$, even on average. Indeed, a better intermediate radius approximation for the average $\Delta \Vobs$ of each galaxy, inside their radial range with observational data, would be $\Delta \Vmod \propto \sqrt r$, as it will be shown in Sec.~\ref{sec:deltaVobs}.

The method considered in this paper is based on the normalized additional velocity (NAV) ($\delta V^2$) as a function of the normalized radius ($\rn$). Namely,
\begin{equation} \label{NAVdef}
   \delta V^2(\rn) \equiv \frac{\Delta V^2(\rn \, r_\mscript{max})}{\Delta V^2(r_\mscript{max})} \,  , 
\end{equation}
where $r_\mscript{max}$ is the largest radial value with RC data and $\rn \equiv r / r_\mscript{max}$ is the normalized galaxy radial coordinate, hence  $\rn \in [0,1]$. 

One can consider other model independent radial values to normalize $\Delta V^2$ (e.g., the stellar disk scale length or the effective galaxy radius $r_\mscript{eff}$), but such options would lead to an intermediate radial point where the $\delta V^2$ curves from different galaxies intersect. Close to this intersection point there is no relevant data to extract since all the curves pass by a such point by the definition. With $r_\mscript{max}$, this intersection happens at the largest radial value, which is a good option to study the intermediate radial behaviour. 

Here we will compare the distributions of $\delta \Vobs$, the observational NAV, and $\delta \Vmod$, the model-inferred NAV.

\subsection{The $\delta\Vobs$ distribution } \label{sec:deltaVobs}

\begin{figure*}
	\begin{tikzpicture}
  		\node (img1)  {\includegraphics[width=0.45\textwidth]{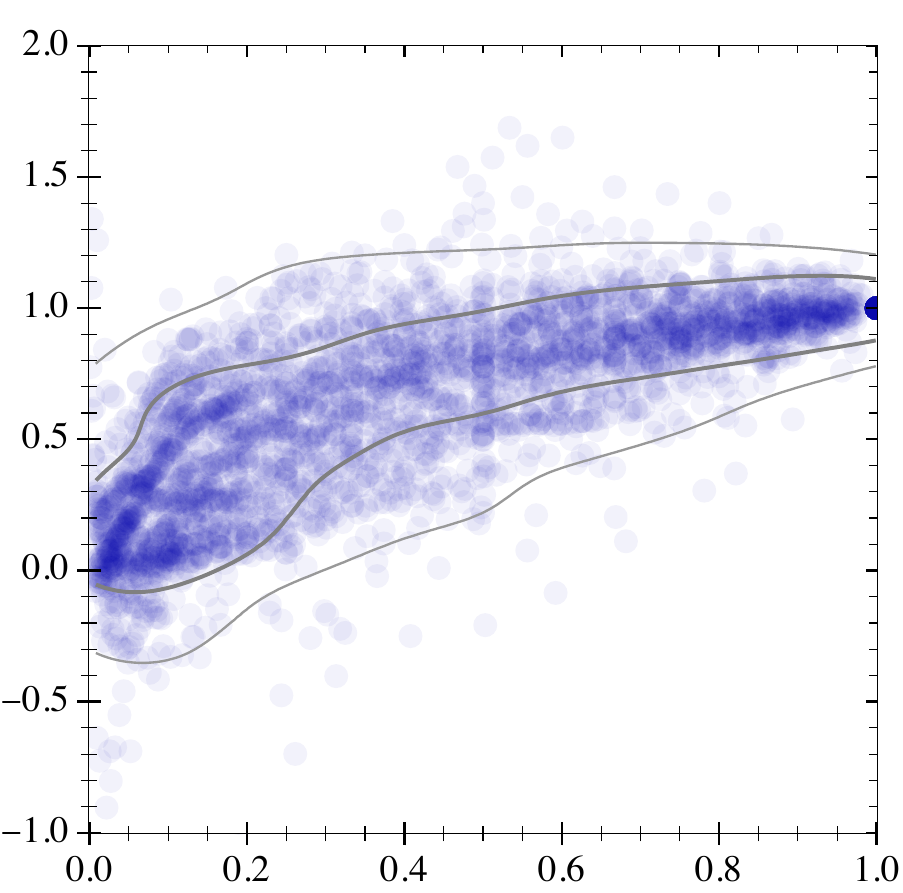}};
		\node[below=of img1, node distance=0cm, yshift=1.1cm, xshift=0.3cm, font=\color{black}] {\large $\rn$};
		\node[left=of img1, node distance=0cm, rotate=90, yshift=-0.9cm, xshift=0.6cm] {\large $\delta \Vobs$};
		\node[above=of img1, node distance=0cm, yshift=-1.5cm, xshift=0.3cm] {\normalsize Includes galaxies with bulge (153 galaxies)};
	\end{tikzpicture} \hspace{-0.3cm}
	\begin{tikzpicture}
  		\node (img2)  {\includegraphics[width=0.45\textwidth]{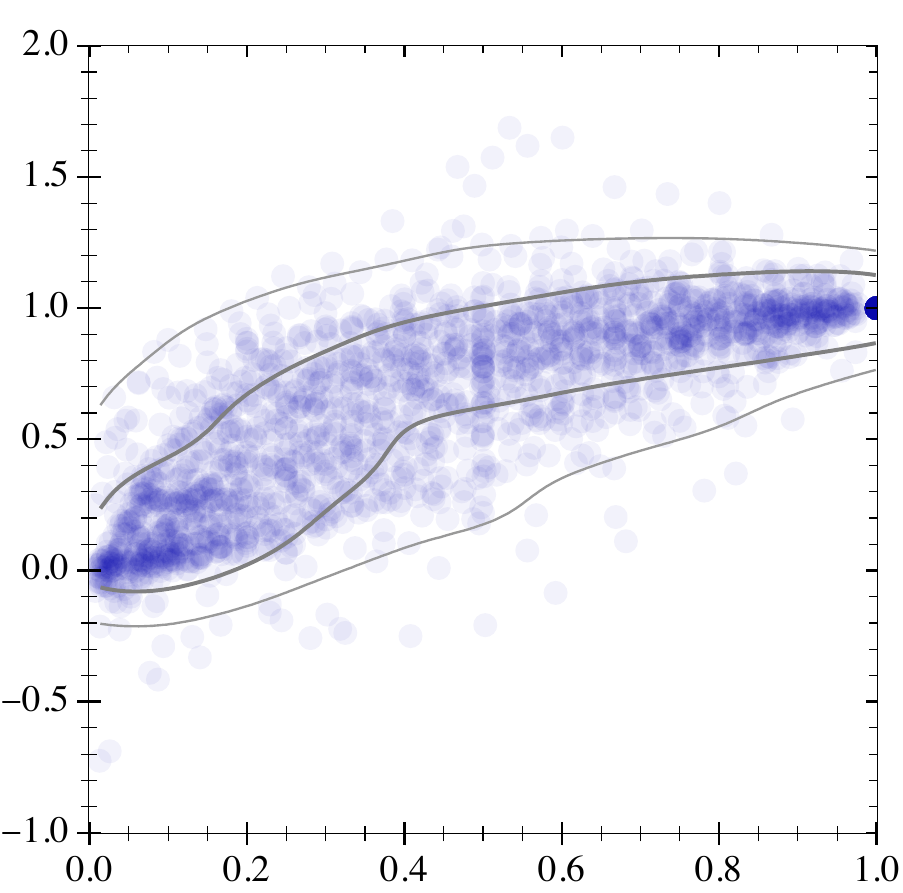}};
		\node[below=of img2, node distance=0cm, yshift=1.1cm, xshift=0.3cm, font=\color{black}] {\large $\rn$};
		\node[above=of img2, node distance=0cm, yshift=-1.5cm, xshift=0.3cm] {\normalsize Excludes galaxies with bulge (122 galaxies)};
	\end{tikzpicture}
	\caption{\textbf{Left plot:} The normalized additional velocity (NAV) plane for 153 galaxies from SPARC with $\YD = 0.5$ and $\YB = 0.6$. Each transparent blue disk is a galaxy data point. The dark blue disk at the point (1, 1) is the superposition of 153 data points: this is the single point that is common to all the galaxies. The grey curves delimit the 1$\sigma$ and 2$\sigma$ highest density regions, which were found from a KDE whose horizontal and vertical bandwidths are (0.068, 0.082), in accordance with the Silverman rule. \textbf{Right plot:} The same as the left plot, but it only includes  galaxies without a dynamically relevant bulge (i.e., 122 galaxies), as classified by SPARC.  The horizontal and vertical bandwidths are (0.072, 0.088). Tabular data on these curves are provided in Table \ref{tab:tabularsigmas}.}
	\label{fig:plotBlueRAR} 
\end{figure*}

Figure \ref{fig:plotBlueRAR} shows the $\delta \Vobs$ distribution due to 153 galaxies from the SPARC catalogue \citep{2016AJ....152..157L}. These are the same 153 galaxies that were used to find the radial acceleration relation (RAR) \citep{McGaugh:2016leg}\footnote{The SPARC catalogue includes 175 galaxies, after eliminating galaxies with low inclinations ($i < 30^\circ$), to reduce the impact of inclination errors, and galaxies that are not sufficiently symmetric (``quality flag 3''), one is left with 153 galaxies. The main RAR plot of \citet{McGaugh:2016leg} also removes data points from some galaxies (those whose relative uncertainty in $V_\mscript{obs}$ is larger than 10\%), this last step is not here considered since its effects for this analysis are negligible: they are larger for $\rn < 0.2$, which is a region that will not be considered. } This figure was derived using the galaxy distance and inclination central values, as provided by SPARC. It is also assumed that the stellar mass-to-light ratios for the disk and the bulge in the 3.6 $\mu$m band from Spitzer are given respectively by $\YD = 0.5$ and $\YB = 0.6$. For the disk, this is exactly the same value used by \citet{McGaugh:2016leg}, and for the bulge we opted for a slight variation (0.6 instead of 0.7), which is in agreement with \citet{Meidt:2014mqa}. This choice is viable and slightly reduces the negative contribution of $\delta \Vobs$ for small radii. Nonetheless, it has no major impact since from the 153 galaxies, 122 have no relevant bulge and since $\rn < 0.2$ will not be considered in this analysis.

\setlength{\tabcolsep}{10pt}
\begin{table*}
\caption{Tabular data for the limiting curves that describe the $1\sigma$ and $2\sigma$ regions in Fig.~\ref{fig:plotBlueRAR}.}
\label{tab:tabularsigmas}
\begin{center}
\begin{tabular}{c cccc c cccc}
\hline
\hline
       & \multicolumn{4}{c}{Sample with bulge (153 galaxies)} & &  \multicolumn{4}{c}{Sample without bulge (122 galaxies)}\\
        \cline{2-5} \cline{7-10}
$r_n$  & 1$\sigma_-$ &  1$\sigma_+$  & 2$\sigma_-$ & 2$\sigma_+$ &  & 1$\sigma_-$ &  1$\sigma_+$  & 2$\sigma_-$ & 2$\sigma_+$  \\ 
  \hline
		0.025 & -0.070 & 0.391 & -0.334 & 0.829 & & -0.072 & 0.281 & -0.208 & 0.674 \\
        0.050 & -0.082 & 0.463 & -0.349 & 0.880 & & -0.080 & 0.348 & -0.212 & 0.750 \\
        0.075 & -0.080 & 0.616 & -0.352 & 0.923 & & -0.079 & 0.393 & -0.212 & 0.812 \\
        0.100 & -0.066 & 0.689 & -0.341 & 0.958 & & -0.070 & 0.432 & -0.205 & 0.872 \\
        0.125 & -0.044 & 0.727 & -0.315 & 0.988 & & -0.055 & 0.475 & -0.193 & 0.923 \\
        0.150 & -0.015 & 0.752 & -0.269 & 1.018 & & -0.035 & 0.532 & -0.177 & 0.963 \\
        0.175 & 0.020 & 0.770 & -0.210 & 1.054 & & -0.009 & 0.608 & -0.158 & 0.997 \\
        0.200 & 0.061 & 0.783 & -0.148 & 1.094 & & 0.022 & 0.672 & -0.136 & 1.028 \\
        0.225 & 0.115 & 0.796 & -0.096 & 1.130 & & 0.059 & 0.721 & -0.111 & 1.054 \\
        0.250 & 0.196 & 0.810 & -0.058 & 1.157 & & 0.105 & 0.764 & -0.084 & 1.079 \\
        0.275 & 0.289 & 0.829 & -0.026 & 1.175 & & 0.163 & 0.802 & -0.055 & 1.099 \\
        0.300 & 0.362 & 0.852 & 0.004 & 1.187 & & 0.225 & 0.835 & -0.026 & 1.117 \\
        0.325 & 0.414 & 0.878 & 0.034 & 1.196 & & 0.286 & 0.868 & 0.003 & 1.133 \\
        0.350 & 0.458 & 0.903 & 0.065 & 1.202 & & 0.350 & 0.898 & 0.031 & 1.148 \\
        0.375 & 0.498 & 0.923 & 0.094 & 1.207 & & 0.439 & 0.924 & 0.059 & 1.164 \\
        0.400 & 0.529 & 0.939 & 0.122 & 1.211 & & 0.530 & 0.946 & 0.086 & 1.181 \\
        0.425 & 0.551 & 0.952 & 0.145 & 1.214 & & 0.570 & 0.964 & 0.110 & 1.197 \\
        0.450 & 0.567 & 0.964 & 0.167 & 1.217 & & 0.593 & 0.979 & 0.131 & 1.213 \\
        0.475 & 0.582 & 0.976 & 0.190 & 1.220 & & 0.608 & 0.994 & 0.152 & 1.227 \\
        0.500 & 0.598 & 0.989 & 0.221 & 1.223 & & 0.622 & 1.007 & 0.175 & 1.236 \\
        0.525 & 0.618 & 1.005 & 0.266 & 1.226 & & 0.635 & 1.020 & 0.207 & 1.244 \\
        0.550 & 0.641 & 1.019 & 0.320 & 1.230 & & 0.648 & 1.034 & 0.254 & 1.249 \\
        0.575 & 0.663 & 1.034 & 0.362 & 1.235 & & 0.663 & 1.047 & 0.309 & 1.254 \\
        0.600 & 0.681 & 1.046 & 0.391 & 1.240 & & 0.678 & 1.060 & 0.352 & 1.258 \\
        0.625 & 0.697 & 1.057 & 0.414 & 1.244 & & 0.691 & 1.072 & 0.383 & 1.261 \\
        0.650 & 0.710 & 1.066 & 0.435 & 1.247 & & 0.705 & 1.083 & 0.410 & 1.263 \\
        0.675 & 0.722 & 1.074 & 0.457 & 1.248 & & 0.717 & 1.093 & 0.433 & 1.265 \\
        0.700 & 0.733 & 1.080 & 0.479 & 1.249 & & 0.728 & 1.102 & 0.457 & 1.266 \\
        0.725 & 0.745 & 1.086 & 0.502 & 1.249 & & 0.740 & 1.110 & 0.478 & 1.267 \\
        0.750 & 0.756 & 1.092 & 0.527 & 1.248 & & 0.751 & 1.116 & 0.499 & 1.267 \\
        0.775 & 0.768 & 1.098 & 0.555 & 1.247 & & 0.762 & 1.122 & 0.522 & 1.266 \\
        0.800 & 0.780 & 1.103 & 0.586 & 1.246 & & 0.773 & 1.128 & 0.549 & 1.264 \\
        0.825 & 0.791 & 1.109 & 0.618 & 1.244 & & 0.784 & 1.132 & 0.581 & 1.262 \\
        0.850 & 0.803 & 1.114 & 0.649 & 1.242 & & 0.795 & 1.135 & 0.615 & 1.258 \\
        0.875 & 0.815 & 1.119 & 0.675 & 1.239 & & 0.806 & 1.139 & 0.648 & 1.254 \\
        0.900 & 0.827 & 1.122 & 0.698 & 1.234 & & 0.818 & 1.140 & 0.675 & 1.249 \\
        0.925 & 0.839 & 1.123 & 0.720 & 1.229 & & 0.829 & 1.141 & 0.700 & 1.243 \\
        0.950 & 0.851 & 1.122 & 0.740 & 1.222 & & 0.841 & 1.139 & 0.723 & 1.236 \\
        0.975 & 0.864 & 1.119 & 0.760 & 1.213 & & 0.853 & 1.134 & 0.745 & 1.228 \\
        1.000 & 0.877 & 1.111 & 0.779 & 1.204 & & 0.867 & 1.126 & 0.765 & 1.219 \\
\hline 
\hline
\end{tabular}
\end{center}
\end{table*}

The observational NAV plane, as displayed in Fig.~\ref{fig:plotBlueRAR}, either including or excluding the galaxies with a stellar bulge, clearly displays a correlation. This correlation is a direct consequence of the need for dark matter or modified gravity. To clarify this point, first we shall consider the case of Newtonian gravity without dark matter. For such case,  $\Vmod = \Vbar$ and therefore $\Delta \Vmod = 0$. Observationally, one would expect $\Delta \Vobs \sim 0$ in this context, hence it should have random oscillations about zero. Therefore, since $\Delta \Vobs(r_\mscript{max})$ would be a random number close to zero, the expected $\delta \Vobs$ should fill out the observational NAV plane (including negative and positive contributions randomly). The latter is clearly not the case for the observational NAV plane inferred from the SPARC data (Fig.~\ref{fig:plotBlueRAR}), hence the displayed correlation demonstrates the need for dark matter or modified gravity.

To proceed with the method, such that it is possible to compare distributions, we need to find a smoothed distribution from the data points. A standard procedure is by using a kernel density estimator (KDE). Whenever we use a KDE in this work, we use a simple and well known one, namely a Gaussian kernel with the Silverman rule for the bandwidths \citep{Silverman_1998, Scott_2014}.  Appendix \ref{app:KDE} covers further KDE details.

Once the KDE for the observational NAV plane is known, one can find the 1$\sigma$ and 2$\sigma$ highest density regions (HDRs). They are regions where most of the $\delta \Vobs$ data points can be found. We stress that they are neither credible nor confidence regions. They show the HDRs contours associated with the probabilities of $68\%$ or $95\%$ respectively. These regions are displayed in Fig.~\ref{fig:plotBlueRAR} and tabulated in Table \ref{tab:tabularsigmas}. 

\begin{figure}
	\begin{tikzpicture}
  		\node (img1)  {\includegraphics[width=0.45\textwidth]{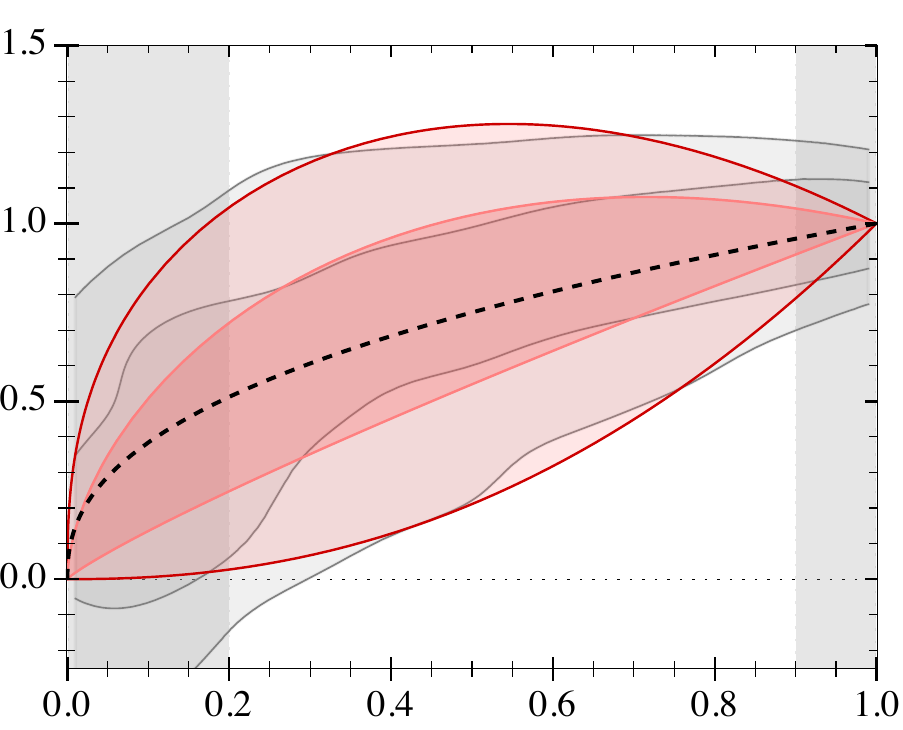}};
		\node[below=of img1, node distance=0cm, yshift=1.1cm, xshift=0.3cm, font=\color{black}] {\large $\rn$};
		\node[left=of img1, node distance=0cm, rotate=90, yshift=-0.9cm, xshift=0.6cm] {\large $\delta V^2$};
	\end{tikzpicture} 
	\caption{Comparison between the 1$\sigma$ and 2$\sigma$ regions, from Fig.~\ref{fig:plotBlueRAR}, with the polynomial models of eq.~\eqref{eq:polyModels}. The vertical shaded regions represent the regions that were not considered for the fits ($\rn < 0.2$ and $\rn > 0.9$), but the plot includes the curves extensions toward all possible $\rn$ values. The dashed black is the curve $(\delta \Vobs)_{\rm best}$, the darker red region is delimited by the curves $(\delta \Vobs)_{1\sigma_\pm}$, while the lighter red by $(\delta \Vobs)_{2\sigma_\pm}$. The mean efficiency \eqref{meanEfficiency} is  $E_\mscript{M} =0.86$.}
	\label{fig:plotPowerLawModel}
\end{figure}

Good models are expected to yield  $\delta \Vmod$ distributions similar to $\delta \Vobs$. A perfect covering of $\delta \Vobs$ is not a necessity because of the presence of observational and KDE errors. About the latter ones, close to $\rn =0$ and $\rn = 1$ there are errors due to the KDE resolution.  We do not consider data closer to the extrema than the inferred bandwidth, in particular we will not consider data in the ranges $\rn < 0.1$ and $\rn > 0.9$. Observational errors on $\delta \Vobs$ close to $\rn\sim 1$ are not relevant, since, by construction, $\delta \Vobs(\rn \sim 1) \sim 1$. On the other hand, for $\rn \sim 0$, $\delta \Vobs$ is particularly sensitive to observational errors, since in this region $\Delta \Vobs \lesssim \Vbar$. By comparing the left and right plots from Fig.~\ref{fig:plotBlueRAR}, which differ by the galaxies with a relevant bulge, the major differences are for $\rn < 0.2$. Due to the observational and KDE errors, we fix the $\rn$ range of especial concern to be $0.2 < \rn < 0.9$.

It is relevant to know simple analytical (polynomial) models that can approximately describe the $\delta \Vobs$ distribution, while also respecting the constraints of passing by the points $(0,0)$ and $(1,1)$ (as realistic models are expected to satisfy). This can be achieved with models of the type $\delta V^2 = \rn^a$ and $\delta V^2 = 2 \rn^b - \rn^c$, where $a, b$ and $c$ are constants. The first case woks for monotonic curves, while the latter for non-monotonic  curves. Within the region $0.2 < \rn < 0.9$, we fit these models considering the following cases: the data from all the galaxies together (denoted by ``best''), or to the curves that describe the upper and lower 1$\sigma$ and 2$\sigma$ limits. The results are
\begin{align}
  &\delta V^2_{\rm best} = \rn^{0.42}\, ,  & \label{eq:polyModels} \\
  &\delta V^2_{1\sigma_-} = \rn^{0.87}\, , \;\;
  &\delta V^2_{1\sigma_+} = 2 \rn^{0.58} - \rn^{1.66}\, ,    \nonumber  \\
  &\delta V^2_{2\sigma_-} = \rn^{2.2}\, ,  
  &\delta V^2_{2\sigma_+} = 2 \rn^{0.38} - \rn^{1.9}\, .  \nonumber
\end{align}
The curves above are plotted in Fig.~\ref{fig:plotPowerLawModel}. Realistic dark matter or modified gravity models are not expect to cover the $\delta \Vobs$ distribution, in the region $0.2 < \rn < 0.9$, better than this polynomial case, which was built exclusively for this purpose.

\subsection{The $\delta \Vmod$ distribution} \label{sec:deltaVmod}

Commonly, one of the parameters that $\Delta \Vmod$ depends on is a global multiplicative constant (e.g., $\rho_s$ for the NFW profile, \citealp{Navarro:1996gj}). Such a parameter, when present, adjusts the magnitude of the dark matter (or modified gravity) contribution. From its definition, $\delta \Vmod$ cannot depend on such parameter. This is a relevant feature about using $\delta \Vmod$: it commonly depends on one less parameter than the original model. 

As a first step for the method application, an analytical expression of $\delta \Vmod$, before any numerical evaluation, can be sufficient as a first comparison with the observational data. This is illustrated in Sec.~\ref{sec:models}.

As a second fast step, one can perform a quick numerical evaluation of the extreme parameter cases in order to verify if the region with observational data can be covered by the model, at least within certain parameter regions. This is specially easy to be done for two-parameter models, which typically become one-parameter models within the $\delta \Vmod$ analysis.

A third test involves determining the best model curves that mimic the 1$\sigma$ and 2$\sigma$ HDRs (as in Fig.~\ref{fig:plotBlueRAR}). The central idea here is to perform a couple of simple fits directly to the sample data, in place of a hundred(s) of galaxy fits. One can test if the model can lead to reasonable approximations to such regions, and, if possible, which parameter values are the most commonly necessary to cover the observational data.

A  general procedure to perform the  fits described above is by using the following integral, 
\begin{equation} \label{Ifit}
	I^{\mscript{($\pm$k)}}(p_i) = \int_0^1 \left (\delta \Vmod(p_i, \rn)  - \lambdaO{$\pm$ k}(\rn) \right )^2 \, \omega(\rn) \,  d\rn \, ,
\end{equation}
where $p_i$ refers to all the model parameters that  $\delta \Vmod$ depends on, $\lambdaO{$\pm$k}(\rn)$ refers to the curve that delimits the k$\sigma$ region of $\Vobs$, with $+k$ corresponding to the upper limit, and $-k$ with the lower limit. The $\lambdaO{$\pm$1}$ and $\lambdaO{$\pm$2}$ functions in tabular form are provided in Table \ref{tab:tabularsigmas}. $\omega(\rn)$ is a weight function. A simple and relevant case, which  we adopt here, is $\omega(\rn) = \Theta(\rn-0.2) \Theta(0.9-\rn)$, where $\Theta$ is the Heaviside theta function. Hence, for the latter case,  only the data in the range $0.2 < \rn < 0.9$ is relevant for the fit.

For some models, the disagreement between $\delta \Vobs$ and $\delta \Vmod$ may be large enough to clearly dismiss them as realistic. As an extreme example, this is the case of Newtonian gravity without dark matter, whose NAV plane would be randomly filled out. From the observational data, it is also easy to dismiss some regions of the radial parameter of either NFW of the Burkert profiles: too low values will generate curves that go much beyond the observational region. A method to quantify the similarity between $\delta \Vobs$ and $\delta \Vmod$ is detailed in Sec.~\ref{sec:ModelEfficiencyIntro}.

For modified gravity approaches, $\delta \Vmod$ typically depends on the 3D baryonic matter distribution of each galaxy, not only the baryonic circular velocities contributions in the plane. It also requires, in general, solving modified Poisson equations.  However, there is a relevant modified gravity model whose $\delta \Vmod$ can be computed promptly and quickly with the available data: this is the case of MOND in its original form \citep{1983ApJ...270..371M}. This is so since $\delta V^2_\mscript{MOND}$ can be written as a function of the baryonic circular velocity Newtonian contribution ($\Vbar$).

\subsection{The NAV efficiency} \label{sec:ModelEfficiencyIntro}

It is convenient to introduce a numerical comparison between the $\Vobs$ and $\Vmod$ distributions. No single number can capture all the data from the 2D plots, but a convenient one, since we are focusing on the 1$\sigma$ and 2$\sigma$ curves, is based on the areas related to these quantities.

First we introduce three areas:
\begin{itemize}
	\item $A_\mscript{obs}(n)$ is the area of the observational NAV region at $n\sigma$ level.
	\item $A_\mscript{mod$\cap$obs}(n)$ is the area of the intersection between the model and observational NAV regions at $n\sigma$ level. Hence, a perfect model is expected to satisfy $A_\mscript{mod $\cap$ obs}(n) = A_\mscript{obs}(n)$.
	\item $A_\mscript{mod$\setminus$obs}(n)$ is the area of the region difference between the model  and the observational NAV regions, both at $n\sigma$ level. A perfect model should yield $A_\mscript{mod$\setminus$obs}(n)=0$.
\end{itemize}

As a first proposal for an efficiency coefficient one may consider $A_\mscript{mod$\cap$obs} / A_\mscript{obs}$. For this case, a perfect model would have an efficiency of 1, while the lowest possible efficiency would be zero, which would correspond to the case in which there is no intersection between the observational and the model NAV regions.  This is interesting, but there should be a penalization for models that predict regions that are outside the observational region.

Considering the above, we define the model efficiency at $n\sigma$ level as
\begin{equation} \label{EnDefinition}
	E_n \equiv \frac{A_\mscript{mod$\cap$obs}(n)  -  A_\mscript{mod$\setminus$obs}(n)}{A_\mscript{obs}(n)} \, .
\end{equation}
Hence, for a given $n$, the highest efficiency value is $E_n = 1$, and $E_n$ can be arbitrarily negative. Any reasonable model should have positive values for $E_n$. Models with negative values are those whose predicted area that agrees with the observational data is smaller than the predicted area that is outside the observational region (at the same $n\sigma$ level).

$E_n$ provides a 1D comparison between the model and observational NAV distributions, which is a simplification with respect to the original 2D data. Since we focus on the $1\sigma$ and $2\sigma$ HDRs, the above model efficiency can be conveniently summarized in a single number as follows, the mean efficiency,
\begin{equation}
	E_\mscript{M} \equiv \frac{E_1 + E_2}{2} \, . \label{meanEfficiency}
\end{equation}
This is a number that we will use to compare the models. To avoid the NAV regions where the observational data are less robust, all the areas are computed in the interval $0.2 < \rn < 0.9$.

\subsection{Individual fits} \label{sec:individual}

Here we compare the NAV results with the standard approach of fitting individually each one of the galaxies. All the individual fits consider the 153 SPARC galaxies.  In the NAV analysis the stellar mass-to-light ratios are kept fixed at their central values. For the individual fits, we consider Gaussian priors on $\logYD \equiv \log_{10} \YD$ and $\logYB \equiv \log_{10}\YB$ centered on $\log_{10} 0.5$ and $\log_{10} 0.6$ respectively, with a variance of $0.1^2$ \citep{Meidt:2014mqa, 2014AJ....148...77M}. This  freedom on the variation of the mass-to-light ratios, which is commonly used within fits of individual galaxies,  has in general relevant impact on the individual galaxies, but the overall sample distribution should not be sensitive to this (and we confirm this here). The NAV method  bypasses some details that are relevant for individual galaxies, but are not relevant for the sample results. 

From the observational data for each galaxy, a $\chi^2$ quantity is defined for a given model with parameters $p_a$ as
\begin{equation}
  \chi^2(p_a) = \sum_i \left(\frac{ V(p_a, r_i) - V_i}{\sigma_i} \right)^2 \, .
\end{equation}
In the above, $V_i$ is the observational circular velocity, with uncertainty $\sigma_i$, and  at the radius $r_i$  (as provided by SPARC). $V(p_a, r_i)$ is the model circular velocity at $r_i$.

Since the priors on $\logYB$ and $\logYD$ are Gaussian, it is possible to implement them by using an effective $\chi^2$, which is defined as follows
\begin{align} \label{chi2eff}
    \chi^2_\mscript{eff}&(p_a, \logYD, \logYB)   \equiv  \\ 
  & \chi^2(p_a) +\left( \frac{\logYD - \log_{10}0.5}{0.1} \right)^2  +
    \left( \frac{\logYB - \log_{10}0.6}{0.1} \right)^2 \, . \nonumber
\end{align}
For galaxies without a bulge, the last term should not be considered. 

The best fit values are here computed from the minimization of $\chi^2_\mscript{eff}$. The minimization of $\chi^2_\mscript{eff}$ leads to the maximum \textit{a posteriori} (MAP) estimation \citep[e.g.,][]{gregory2010bayesian}. See also the supplementary information section of \citep{Rodrigues:2018duc}.

Since some galaxies display multiple local minima, we performed several minimizations for each galaxy (at least eight), with different starting points and minimization meta-parameters. They were performed using the code \texttt{MAGMA}\footnote{\href{https://github.com/davi-rodrigues/MAGMA}{github.com/davi-rodrigues/MAGMA}.} \citep{Rodrigues:2018duc}, whose implementation is based on the differential evolution optimization algorithm  \citep{9783540313069}. To find a best fit, i.e. a global minimum, global optimization techniques, as the differential evolution method, are more efficient than MCMC-based procedures, this since that is the single purpose of the former.

The results of the individual fits can be found in \ref{app:IndividualFits}.

\section{The models and the results} \label{sec:models}

\subsection{Arctan$_\alpha$ model}

\subsubsection{Model definition}

This is a toy model whose purpose is to clarify the method and the relation between other models results.  This model is a generalization of a model studied by \citet{Courteau:1997wu}. It is defined directly from the dark matter circular velocity, that is, it was not defined from a density profile. We generalize the original model by considering
\begin{equation} \label{arctanModel}
  V_{\rm arctan}(r) = V_\mscript{c} \left(\frac {2}{\pi} \right)^\alpha  \arctan^\alpha \left( \frac {r}{r_\mscript{t}}\right ) \, ,
\end{equation}
where $V_\mscript{c}$, $r_\mscript{t}$ and $\alpha$ are constants. The two first constants may change from galaxy to galaxy. The original case of \citet{Courteau:1997wu} corresponds to $\alpha=1$. 

\subsubsection{NAV analysis} \label{sec:arctanNAV}

Since $V_{\rm arctan}^2$ is the dark matter contribution to the circular velocity, then, from eq.~\eqref{DVobs}, $\Delta V^2_{\rm arctan}(r) = V^2_{\rm arctan}(r)$. Hence, using eq.~\eqref{NAVdef},
\begin{equation}
  \delta V^2_{\rm arctan}(\rn) = \arctan^{2 \alpha} \left (\frac {\rn}{r_\mscript{tn}}\right )  \bigg / \arctan^{2 \alpha} \left( r^{-1}_\mscript{tn}\right ) \, ,
\end{equation}
with $r_\mscript{tn} = r_\mscript{t}/r_\mscript{max}$. The space of parameters was reduced from three free parameters, in eq.~\eqref{arctanModel}, to two: $r_\mscript{tn}$ and $\alpha$.

For a given $\rn$ value ($0 < \rn < 1$), one can analytically compute the cases of very large or very small $r_\mscript{tn} $ values,

\begin{equation} \label{arctanExtrema}
  \delta V^2_{\rm arctan} = 
  \begin{cases}
    1  & \mbox{, for small $r_\mscript{tn}$.}\\[.3cm]
    \rn^{2 \alpha}  &\mbox{, for large $r_\mscript{tn}$.}      
  \end{cases}
\end{equation}

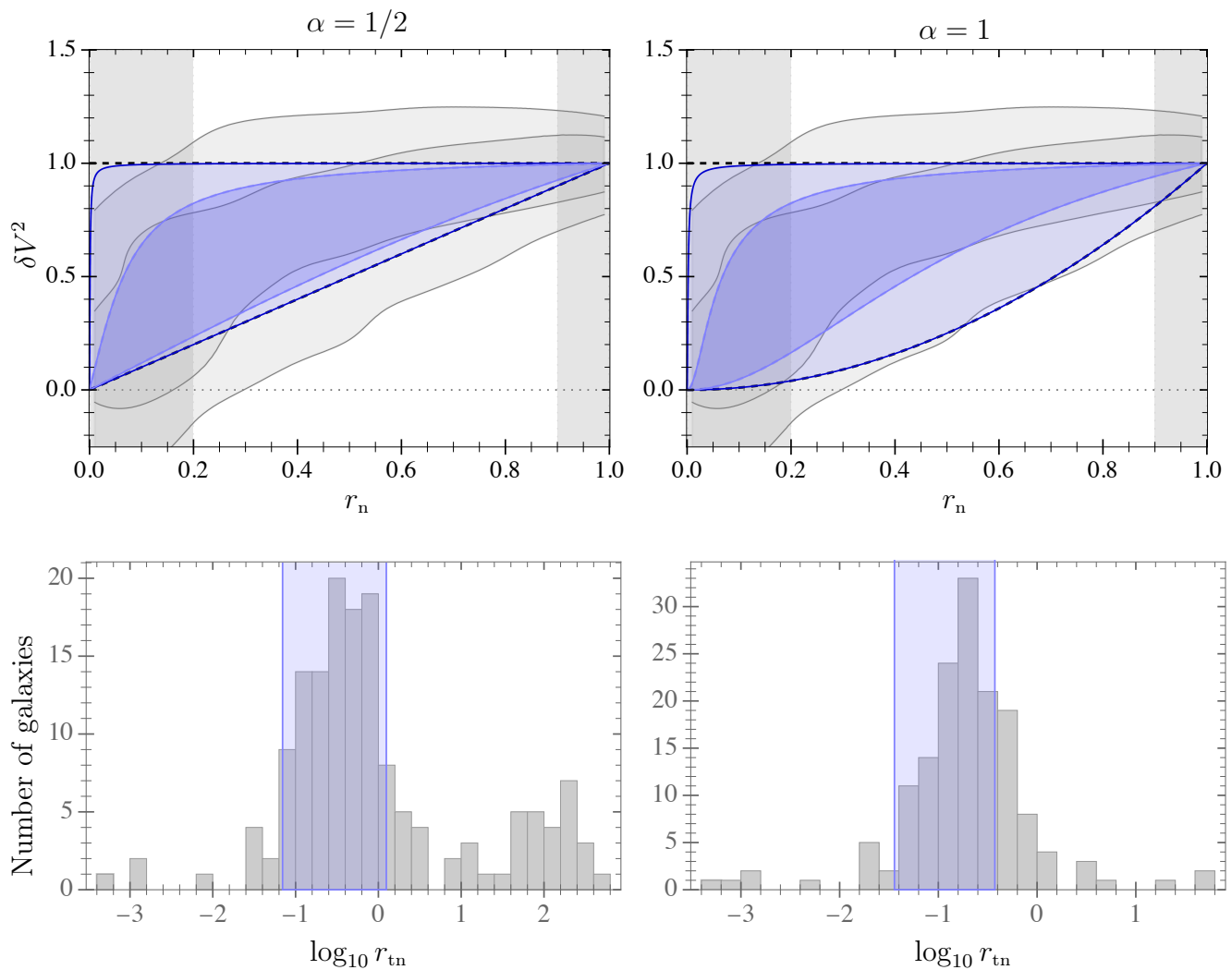
\begin{figure*}
	\begin{tikzpicture}
  		\node (img2)  {\includegraphics[width=0.45\textwidth]{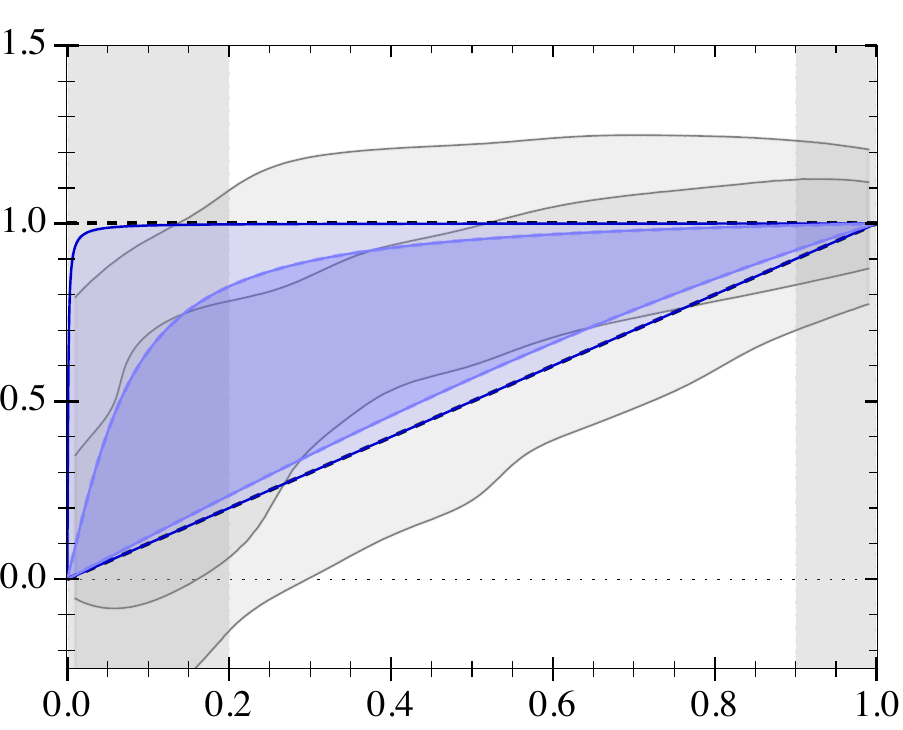}};
		\node[below=of img2, node distance=0cm, yshift=1.1cm, xshift=0.3cm, font=\color{black}] {\large $\rn$};
		\node[left=of img1, node distance=0cm, rotate=90, yshift=-0.8cm, xshift=0.6cm] {\large $\delta V^2$};
		\node[above=of img2, node distance=0cm, yshift=-1.5cm, xshift=0.3cm] {\large $\alpha = 1/2$};
	\end{tikzpicture}\hspace{-0.1cm}
	\begin{tikzpicture}
  		\node (img1)  {\includegraphics[width=0.45\textwidth]{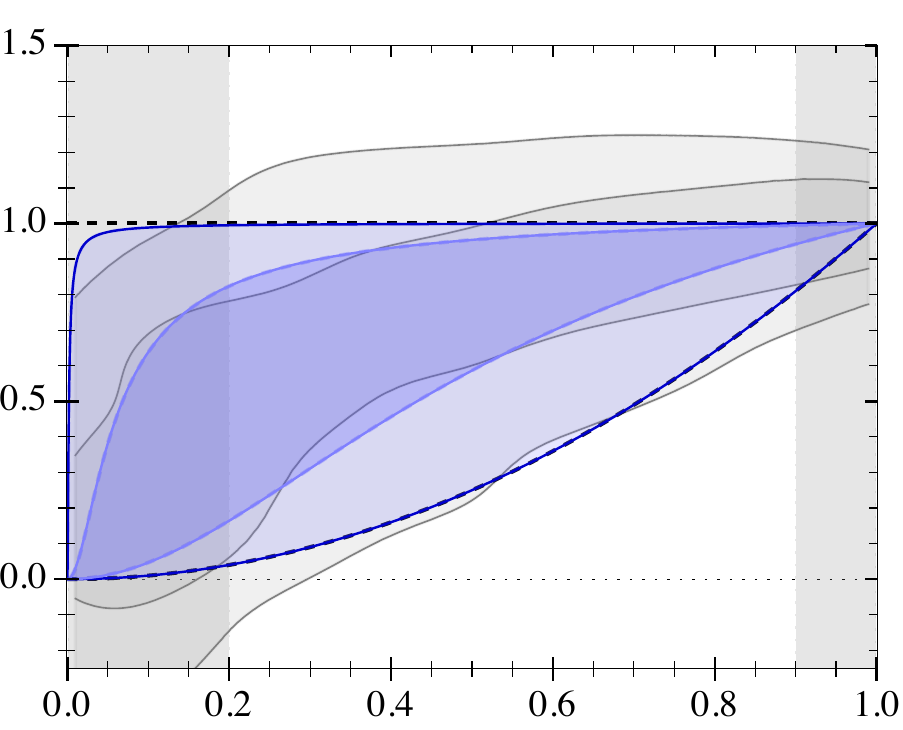}};
		\node[below=of img1, node distance=0cm, yshift=1.1cm, xshift=0.3cm, font=\color{black}] {\large $\rn$};
		\node[above=of img1, node distance=0cm, yshift=-1.5cm, xshift=0.3cm] {\large $\alpha = 1$};
	\end{tikzpicture} \\
	\begin{tikzpicture}
  		\node (img2)  {\includegraphics[width=0.45\textwidth]{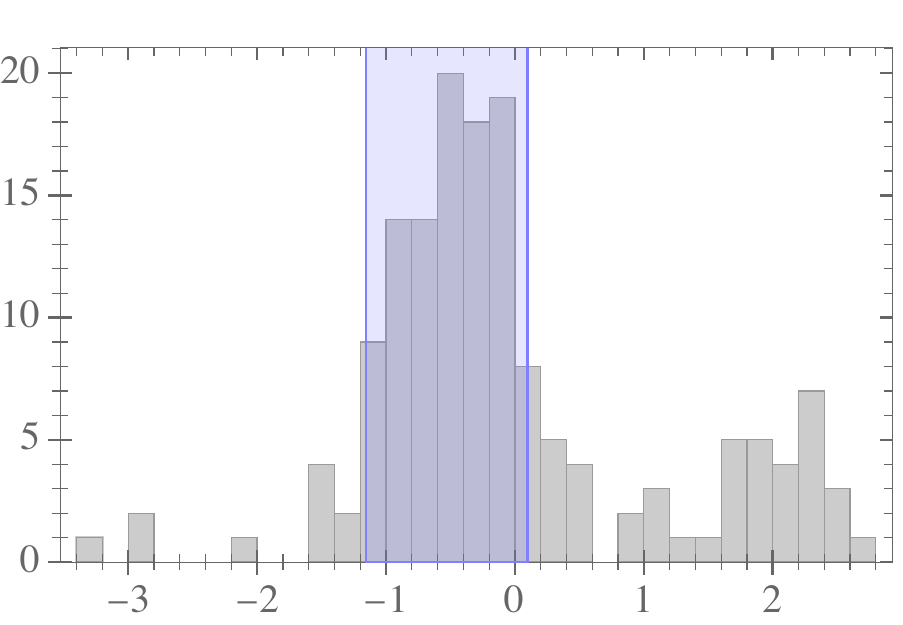}};
		\node[below=of img2, node distance=0cm, yshift=1.1cm, xshift=0.3cm, font=\color{black}] {\large $\log_{10} r_\mscript{tn}$};
		\node[left=of img1, node distance=0cm, rotate=90, yshift=-0.8cm, xshift=1.5cm] {\large Number of galaxies};
	\end{tikzpicture}
	\begin{tikzpicture}
  		\node (img1)  {\includegraphics[width=0.45\textwidth]{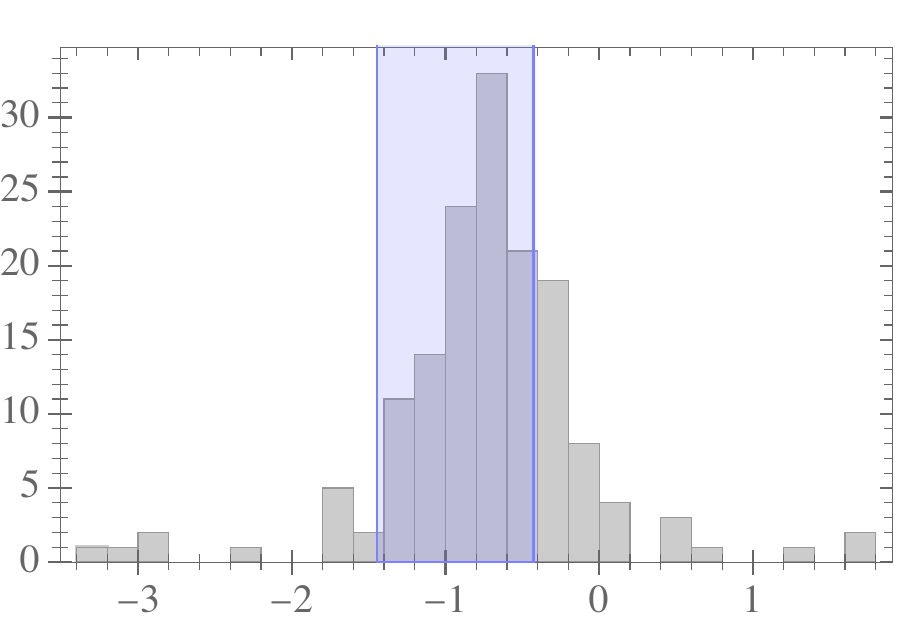}};
		\node[below=of img1, node distance=0cm, yshift=1.1cm, xshift=0.3cm, font=\color{black}] {\large $\log_{10} r_\mscript{tn}$};
	\end{tikzpicture} \hspace{-0.3cm}
	\caption{
	 \textbf{Top.} The NAV plane for the Arctan model with $\alpha = 1/2$ (left) and $\alpha = 1$ (right). See also Fig.~\ref{fig:plotPowerLawModel}. The two black dashed lines show the models extremum cases, that is $r_{\mscript{tn}} \to \infty$ (lower limit in the plot) or $r_{\mscript{tn}} \to 0$ (upper limit in the plot), see eq.~\eqref{arctanExtrema}. The two bluish regions are the arctan models best approximations for the observational 1$\sigma$ and 2$\sigma$ regions (\ref{rtnAlpha1}-\ref{rtnAlpha12}). The mean efficiency \eqref{meanEfficiency} of each case is respectively 0.70 and 0.60. 
	 \textbf{Bottom.} Histograms on the distribution of $r_\mscript{tn}$ from the individual fits,  for the 153 SPARC galaxies. The case with $\alpha = 1/2$ has a strong tendency of generating very large $r_\mscript{tn}$ values, several of them are dynamically equivalent to the $r_\mscript{tn} \to \infty$ case. The transparent blue rectangle depicts the 1$\sigma$ region of the corresponding $\alpha$ case, as inferred from eqs.~(\ref{rtnAlpha1}-\ref{rtnAlpha12}). There are no data beyond the plotted regions in these histograms.
	 }
	\label{fig:plotArctan} 
\end{figure*}

Comparing with Fig.~\ref{fig:plotPowerLawModel}, one can conclude that the upper limit is too strong, it leaves a region without proper covering. No value of $\alpha$ can change that. The lower bound in the NAV plane depends on the $\alpha$ value. Together with eq.~\eqref{eq:polyModels}, $\alpha$ can be chosen to improve concordance with the data. For instance, choosing $\alpha = 1$ (or $\alpha = 1.1$) seems to be a good choice: large $r_\mscript{tn}$ values will be responsible for the lower bound in the NAV plane. On the other hand, the choice $\alpha = 1/2$, which leads to $\delta V_\mscript{arctan}^2 \sim \rn$, covers significantly less observational data than the $\alpha = 1$ case. Indeed, these limiting cases are plotted in Fig.~\ref{fig:plotArctan}.

After considering the analytical and numerical evaluations of the extreme cases, one can proceed to the third step as described in Sec.~\ref{sec:deltaVmod}. We use the integral \eqref{Ifit} in the interval $0.2 < \rn < 0.9$ to find the best approximations for the observational 1$\sigma$ and 2$\sigma$ regions. The results are
\begin{align}
	&1\sigma: \;  0.04 < r_\mscript{tn} < 0.37 \, , \label{rtnAlpha1}\\
	&2\sigma: \; 0 < r_\mscript{tn} < \infty \, ,\nonumber
\end{align}
for the case $\alpha = 1$, and
\begin{align}
	&1\sigma: \;  0.07 < r_\mscript{tn} < 1.25 \, , \label{rtnAlpha12}\\
	&2\sigma: \; 0 < r_\mscript{tn} < \infty \, , \nonumber
\end{align}
for the case $\alpha = 1/2$. The corresponding curves are also plotted in Fig.~\ref{fig:plotArctan}.

At last, we compute the NAV efficiency, as defined in Sec.~\ref{sec:ModelEfficiencyIntro}, for $\alpha = 1/2$,
\begin{align}
    &E_1 = 0.71  \; \;  [0.782 - 0.072] \, ,\nonumber \\
    &E_2 = 0.49  \; \;  [0.489 - 0.000] \, .
\end{align}
The numbers in square brackets refer to the positive and negative contributions from eq.~\eqref{EnDefinition}, they display an intermediate step in the computation.

For $\alpha = 1$, 
\begin{align}
    &E_1 = 0.69  \; \;  [0.764 - 0.070] \, , \nonumber \\
    &E_2 = 0.70  \; \;  [0.709 - 0.005] \, .
\end{align}

In comparison with $\alpha = 1$, the $\alpha = 1/2$ case lacks data diversity, leading to its efficiency being penalized. Consequently, the individual fits using this model will lead to larger $\chi^2$ values, as it will be shown. 

In summary, the mean efficiencies for $\alpha =1/2$ and $\alpha = 1$ read respectively, 
\begin{equation}
  E_\mscript{M}^{(1/2)} = 0.60 \, , \; E_\mscript{M}^{(1)} = 0.70
\end{equation}

\subsubsection{Comparison with the individual fits results} \label{sec:arctanIndividual}

The individual fits are performed considering the following parameters as free: $V_\mscript{c}, r_\mscript{t}, \YD$ and, when a bulge is present, $\YB$. More specifically, the quantities $r_\mscript{t}$ and $V_\mscript{c}$ are fitted with flat priors, but they are restricted to the following ranges, to ensure reasonable values for these quantities and improve computational speed: $0.05 < r_\mscript{t} \, ( {\rm kpc}) < 500$ and $10 < V_\mscript{c} \, \left( \frac{\rm km}{\rm s} \right) < 1000$. $\YD$ and $\YB$ are associated with Gaussian priors, as detailed in the previous section.

Figure \ref{fig:plotArctan} also shows the distribution of $r_\mscript{tn}$ values as found from the individual fits. As expected from the previous section, for $\alpha = 1/2$ several galaxies lead to large  $r_\mscript{tn}$ values, essentially $r_\mscript{tn}\to \infty$. Also, as expected, the previous analysis can be used to infer a range of $r_\mscript{tn}$ that is  the most common within the individual fits. 

It was shown that the mean NAV efficiency $(E_\mscript{M})$ for the $\alpha = 1$ case is larger than the $\alpha = 1/2$ case. Consequently, one should expect that the case $\alpha = 1$ will lead to typically lower $\chi^2$ values. Indeed, comparing the median $\chi^2$ values for each case we find that
\begin{equation}
  \widetilde{\chi^2_{\mscript{arctan}_{1/2}}}\approx 18 \; \mbox{ and } \;   \widetilde{\chi^2_{\mscript{arctan}_{1}}} \approx 9 \, .
\end{equation}
We use a tilde to denote the median. The mean $\chi^2$ also favors the $\alpha = 1$ case, however it is a less robust sample description, since it is prone to large changes due to the data of a single galaxy. In Sec.~\ref{sec:modelComparisons} we compile and compare all the models results.

\subsection{NFW} \label{sec:NFW}

\subsubsection{Model definition} \label{sec:modelNFW}
The Navarro-Frenk-White (NFW) dark matter profile was derived from N-body simulations for a dark matter-only universe \citep{Navarro:1996gj, 0521857937}. It has achieved great success in several areas, it is commonly considered to be the standard dark matter halo within several astrophysical applications. However, since it does not include the influence of baryons, it is not unexpected  that systematical deviations can be found in some systems, including galaxies \citep[e.g.,][]{deBlok:2009sp, Governato:2012fa,  DiCintio:2014xia, Oman:2015xda, DelPopolo:2016emo, Dutton:2019gor, 2022MNRAS.514.3510F}. Baryonic physics should be able to provide certain degree of diversity to dark matter halos that is beyond the diversity that can be found from pure dark matter simulations. This includes the cusp/core controversy and goes beyond it \citep{Oman:2015xda}.

The NFW profile is a two-parameter  cuspy dark matter halo  whose density profile is given by 
\begin{equation}
	\rho_\mscript{NFW}(r) = \frac{\rho_\mscript{s}}{ r/r_\mscript{s}(1 + r/r_\mscript{s})^2}\, . \label{nfwrho}
\end{equation}
In the above, $r$ is the spherical radial coordinate, while $r_\mscript{s}$  and $\rho_\mscript{s}$ are constants that can change from galaxy to galaxy. 

The internal mass of this profile is given by
\begin{equation}
  M_\mscript{NFW}(r) = 4 \pi r^3 \rho_s \left [ \ln \left ( \frac{r + r_\mscript{s}}{r_\mscript{s}}\right ) - \frac{r}{r + r_\mscript{s}}\right ] \, .
\end{equation}

\subsubsection{NAV analysis} \label{sec:NAVNFW}
The additional velocity is the dark matter contribution, hence 
\begin{equation}
  \Delta V^2_\mscript{NFW}(r) = G \frac{M_\mscript{NFW}(r)}{r} \, .
\end{equation}
Using the $M_\mscript{NFW}$ expression,
\begin{align}
  \delta V^2_\mscript{NFW}(r_n) & =  \frac{\Delta V^2_\mscript{NFW}(r_n \, r_\mscript{max})} {\Delta V^2_\mscript{NFW}(r_\mscript{max})} \nonumber \\[.3cm] 
  & = \frac{1}{\rn}\frac{\frac{1}{1 + r_\mscript{sn}/\rn} - \ln ({1 + \rn/r_\mscript{sn}})}{\frac{1}{1 + r_\mscript{sn}} - \ln ({1 + 1/r_\mscript{sn}})} \, .
\end{align}

For a given $0<\rn<1$, one can compute the asymptotic limits,
\begin{equation} \label{nfwExtrema}
  \delta V^2_\mscript{NFW} = 
  \begin{cases}
    \rn^{-1}  & \mbox{, for small $r_\mscript{sn}$.}\\[.3cm]
    \rn  &\mbox{, for large $r_\mscript{sn}$.}      
  \end{cases}
\end{equation}

The above result for small $r_\mscript{sn}$ does not contain the tight bound that the Arctan model has for small $r_\mscript{tn}$ and for any $\alpha$. On the other hand, the lower bound in the NAV plane, given by $\rn$, is too stringent, and it is the same of the Arctan model with $\alpha = 1/2$.

\begin{figure*}
	\begin{tikzpicture}
  		\node (img1)  {\includegraphics[width=0.45\textwidth]{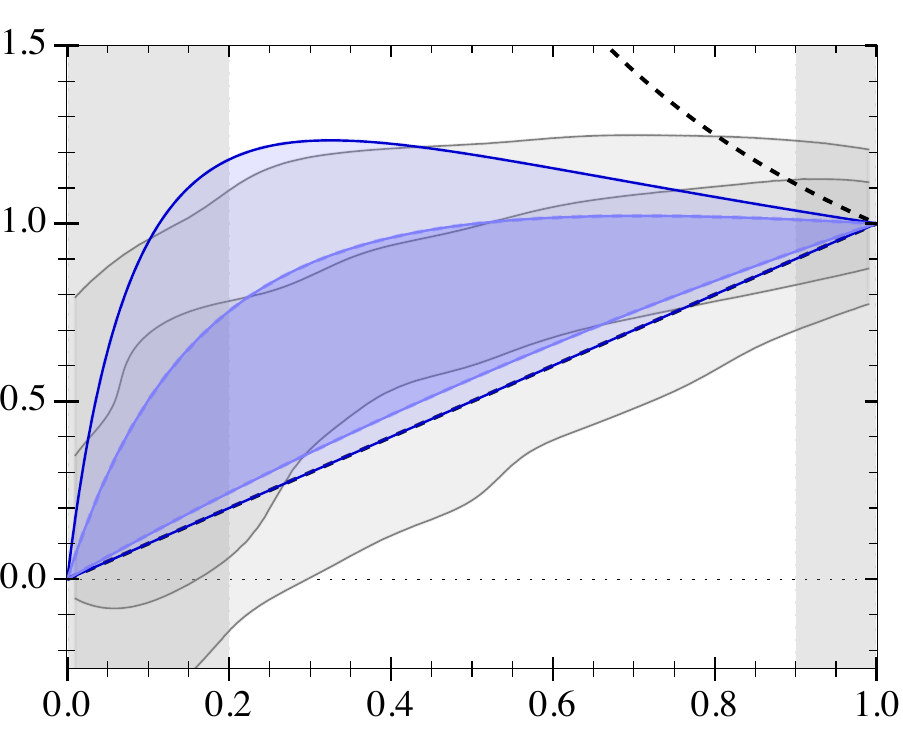}};
		\node[below=of img1, node distance=0cm, yshift=1.1cm, xshift=0.3cm, font=\color{black}] {\large $\rn$};
		\node[left=of img1, node distance=0cm, rotate=90, yshift=-0.8cm, xshift=0.6cm] {\large $\delta V^2$};
		\node[above=of img1, node distance=0cm, yshift=-1.5cm, xshift=0.3cm] {\large NFW};
	\end{tikzpicture} \hspace{-0.3cm} 
	\begin{tikzpicture}
  		\node (img1)  {\includegraphics[width=0.45\textwidth]{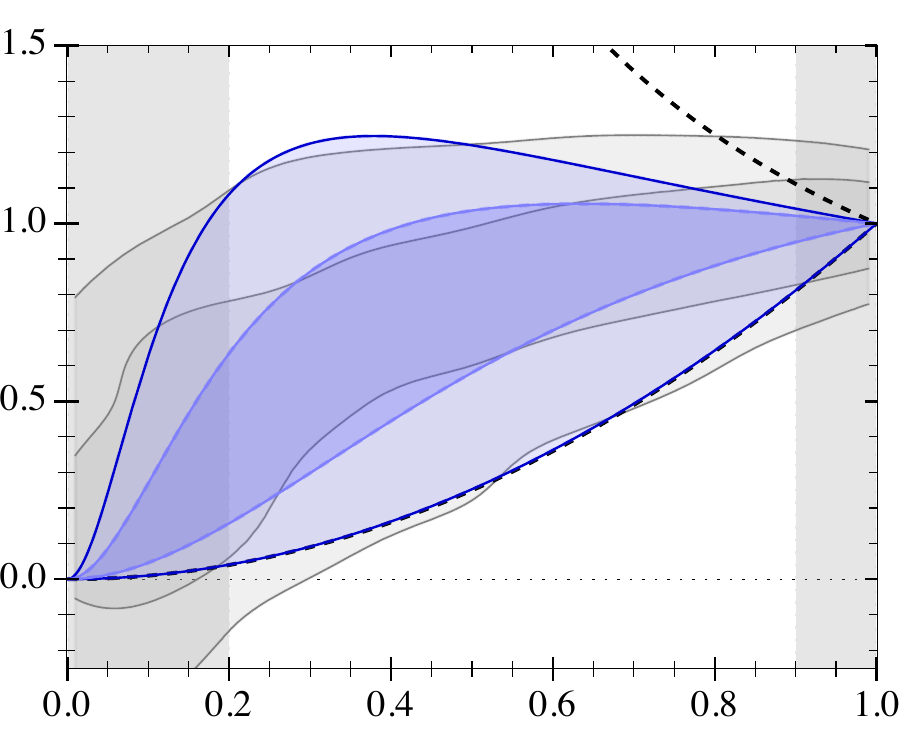}};
		\node[below=of img1, node distance=0cm, yshift=1.1cm, xshift=0.3cm, font=\color{black}] {\large $\rn$};
		\node[above=of img1, node distance=0cm, yshift=-1.5cm, xshift=0.3cm] {\large Burkert};
	\end{tikzpicture}\\
	\begin{tikzpicture}
  		\node (img1)  {\includegraphics[width=0.45\textwidth]{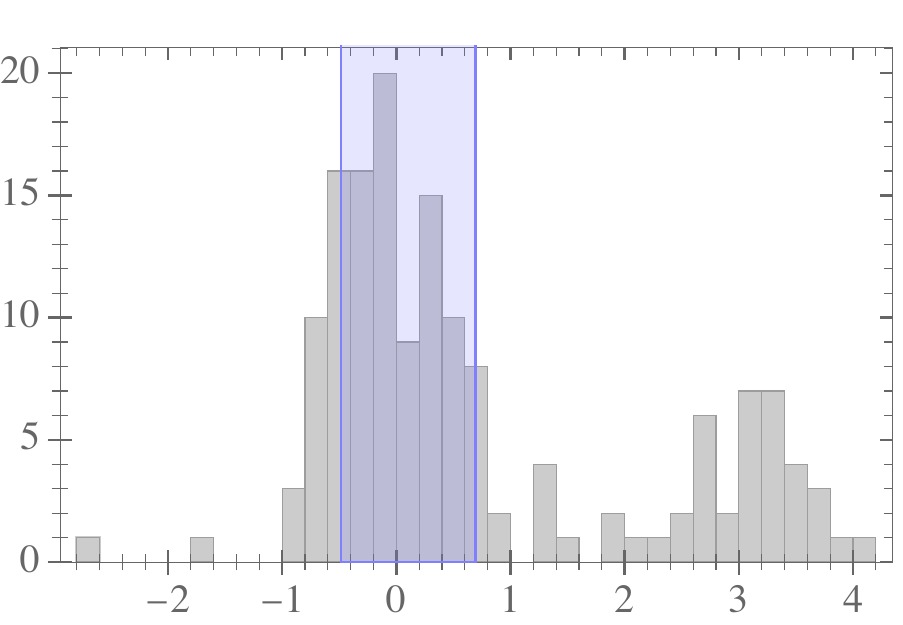}};
		\node[below=of img1, node distance=0cm, yshift=1.1cm, xshift=0.3cm, font=\color{black}] {\large $\log_{10} r_\mscript{sn}$};
		\node[left=of img1, node distance=0cm, rotate=90, yshift=-0.8cm, xshift=1.5cm] {\large Number of galaxies};
	\end{tikzpicture} \hspace{-0.3cm} 
	\begin{tikzpicture}
  		\node (img1)  {\includegraphics[width=0.45\textwidth]{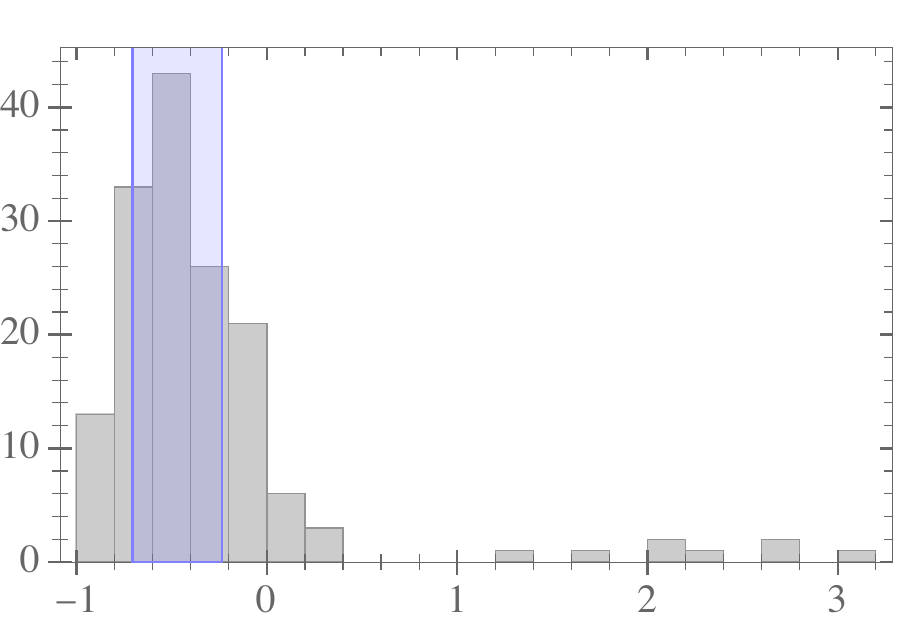}};
		\node[below=of img1, node distance=0cm, yshift=1.1cm, xshift=0.3cm, font=\color{black}] {\large $\log_{10} r_\mscript{cn}$};
	\end{tikzpicture} 
	\caption{
	 \textbf{Top.} The NAV plane for the NFW (left) and Burkert (right) models. See also Figs.~\ref{fig:plotPowerLawModel}, \ref{fig:plotArctan}. The two black dashed lines show the models extremum cases, see eq.~\eqref{nfwExtrema} and eq.\eqref{burkertExtrema} respectively. The two bluish regions are the NFW or Burkert models best approximations for the observational 1$\sigma$ and 2$\sigma$ regions \eqref{rsnHDRLimits}, \eqref{rcnHDRLimits}. The mean efficiency values are 0.70 and 0.80 respectively, from eqs.~(\ref{meanEfficiencyNFW}, \ref{meanEfficiencyBurkert}).
	 \textbf{Bottom.} Histograms on the distribution of $r_\mscript{sn}$ and $r_\mscript{cn}$ from the individual NFW and Burkert fits. The transparent blue rectangle displays the 1$\sigma$ region as inferred from eq.~\eqref{rsnHDRLimits} or eq.\eqref{rcnHDRLimits}. Several galaxies favor very large $r_\mscript{sn}$ values (with $r_\mscript{s} > 10 r_\mscript{max}$). For Burkert, only a few galaxies favor $r_\mscript{c} > 10 r_\mscript{max}$. Both histograms display the complete data, there is no galaxy data beyond the plotted regions.
	 }
	\label{fig:plotNFWburkert} 
\end{figure*}
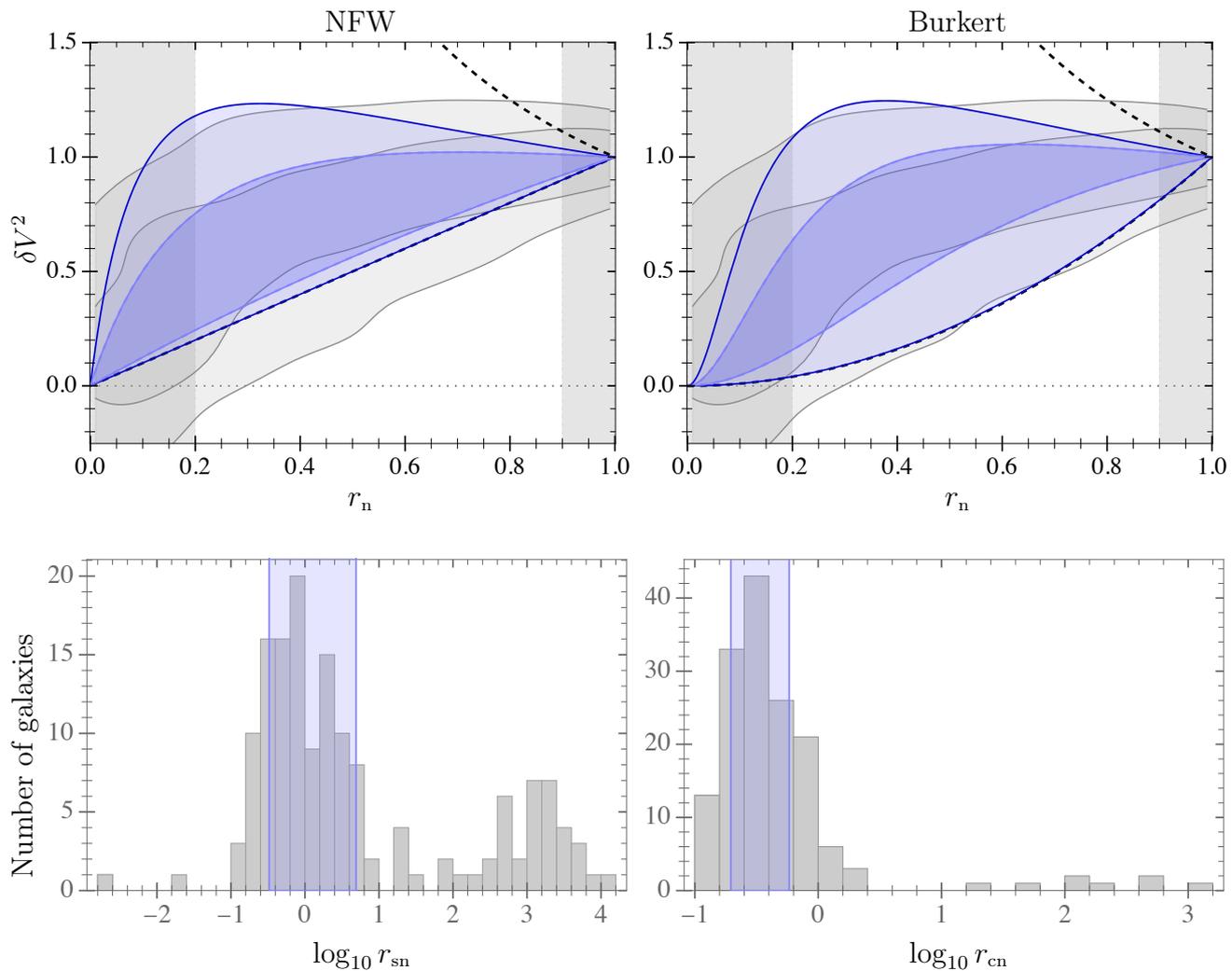

We proceed now to fit the 1$\sigma$ and 2$\sigma$ regions using the integral \eqref{Ifit} in the interval $0.2 < \rn < 0.9$. The results are
\begin{align}
	&1\sigma: \;  0.33 < r_\mscript{sn} < 4.9 \, , \nonumber \\
	&2\sigma: \;  0.15 < r_\mscript{sn} < \infty \, .\label{rsnHDRLimits}
\end{align}
These regions are plotted in Fig.~\ref{fig:plotNFWburkert}.

The NAV efficiency for the NFW model is computed as
\begin{align}
    &E_1 = 0.77  \; \;  [0.843 - 0.072] \, , \nonumber \\
    &E_2 = 0.63  \; \;  [0.648 - 0.015] \, .
\end{align}
Hence, the mean efficiency reads,
\begin{equation}
  E_\mscript{M} = 0.70\,  . \label{meanEfficiencyNFW}
\end{equation}
 The NFW model has the same mean efficiency of the arctan model with $\alpha=1$, and it is better than the $\alpha = 1/2$ case due to the upper part of the NAV plane.

\subsubsection{Comparison with the individual fits results} \label{NFWindividual}

The individual NFW fits are done following the same procedures detailed in Sec.~\ref{sec:arctanIndividual}, with the exception for the NFW parameters constraints, which read: $\rho_\mscript{s}  > 10 \, {\rm M}_\odot / {\rm kpc}^3 $ and $0.05 < r_\mscript{s} ({\rm kpc}) < 10^4$. 

Figure \ref{fig:plotNFWburkert} shows the distribution of $r_\mscript{sn}$ values as found from the individual fits. As expected from the previous section, several galaxies imply very large values for $r_\mscript{sn}$. Since the limit $r_\mscript{sn} \to \infty$ approaches the asymptotic value logarithmically, the NFW model requires larger $r_\mscript{sn}$ than the arctan model with $\alpha = 1/2$. Also, as expected, the NAV analysis can be used to infer an approximate range of $r_\mscript{sn}$ values that are most common within the individual fits. 

The median $\chi^2$ value for NFW reads $\widetilde{\chi^2_\mscript{NFW}} \approx 15$. This is  better than the median $\chi^2$ value of the arctan model with $\alpha = 1/2$, and  worse than the median $\chi^2$ for the $\alpha = 1$ case. Further model comparisons will be done in Sec.~\ref{sec:modelComparisons}.

\subsection{Burkert profile} \label{sec:ModelBurkert}

\subsubsection{Model definition}

The Burkert profile \citep{1995ApJ...447L..25B} is a two-parameter cored dark matter halo that has been extensively studied and which has achieved phenomenological success \citep[e.g.,][]{Salucci:2000ps, Salucci:2007tm, Rodrigues:2017vto, Li:2020iib}. Its density profile is given by 
\begin{equation}
	\rho_\mscript{Bur}(r) = \frac{\rho_\mscript{c}}{(1 + r/r_\mscript{c})(1 + r^2/r_\mscript{c}^2)}\, . \label{burkertrho}
\end{equation}
In the above, $r$ is the spherical radial coordinate, while $r_\mscript{c}$  and $\rho_\mscript{c}$ are constants that can change from galaxy to galaxy. 

The internal mass of the Burkert profile reads,
\begin{align}
	M_\mscript{Bur}(r) & = 4 \pi \int_0^r \rho_\mscript{Bur}(r) r^2 \, dr \nonumber \\
	& = 2 \pi \, \rho_\mscript{c} r_\mscript{c}^3 \, \xi\left(\frac{r}{r_\mscript{c}}\right) \, , \label{burkertmass}
\end{align}
where
\begin{equation}
	\xi(x) \equiv \ln\left((1+x) \sqrt{1+x^2} \right) - \tan^{-1}(x)  \, .
\end{equation}

\subsubsection{NAV analysis} \label{sec:NAVBurkert}

It is convenient to introduce the normalized core radius as
\begin{equation}
	 r_\mscript{cn} \equiv	\frac{r_\mscript{c}}{r_\mscript{max}}\, .
\end{equation}

Using that, for a spherical mass distribution, $V^2(r) = G M(r)/r$, we can now compute $\Delta \Vmod$ and $\delta \Vmod$ as
\begin{align}
	\Delta V^2_\mscript{Bur}(\rn) &= 2 \pi  \frac{G \rho_\mscript{c} r_\mscript{c}^3}{r_\mscript{max} } \frac{1}{\rn} \xi\left(\frac{\rn}{r_\mscript{cn}}\right), \label{burkertDeltaV2}\\[.2cm]
	\delta V^2_\mscript{Bur}(\rn) &=\frac{1}{\rn} \frac{\xi(\rn/r_\mscript{cn})}{\xi(1/r_\mscript{cn})} . \label{burkertdeltav2}
\end{align} 

For given $0<\rn<1$, one can compute the asymptotic limits,
\begin{equation} \label{burkertExtrema}
  \delta V^2_\mscript{Bur} = 
  \begin{cases}
    \rn^{-1}  & \mbox{, for small $r_\mscript{cn}$.}\\[.3cm]
    \rn^2  &\mbox{, for large $r_\mscript{cn}$.}      
  \end{cases}
\end{equation}

This model has the same asymptotic for small $r_\mscript{cn}$ as the NFW model, but for large $r_\mscript{cn}$ it behaves like the arctan model with $\alpha = 1$. Hence, it is expect that its NAV efficiency is larger than the previous models, and this will be the case.

We fit the 1$\sigma$ and 2$\sigma$ regions using the integral \eqref{Ifit} in the interval $0.2 < \rn < 0.9$. The results are
\begin{align}
	&1\sigma: \;  0.20 < r_\mscript{cn}  < 0.58 \, , \nonumber \\
	&2\sigma: \;  0.12 < r_\mscript{cn}  < 35 \, .\label{rcnHDRLimits}
\end{align}
These regions are plotted in Fig.~\ref{fig:plotNFWburkert}.

The efficiency for the Burkert model reads
\begin{equation}
  E_1 = 0.74 \, \mbox{ and }  E_2 = 0.86 \, \mbox{, hence } E_\mscript{M} = 0.80\,  . \label{meanEfficiencyBurkert}
\end{equation}
This is the highest $E_\mscript{M}$ value for the models considered up to this point, and very close to the $E_\mscript{M}$ value of the polynomial description in Fig.~\ref{fig:plotPowerLawModel}.

\subsubsection{Comparison with the individual fits results} \label{sec:Burkertindividual}

The individual Burkert fits are done following the same procedures detailed in Sec.~\ref{sec:arctanIndividual}, with the exception for the Burkert parameter constraints, which read: $\rho_\mscript{c}  > 10 \, {\rm M}_\odot / {\rm kpc}^3 $ and $0.01 < r_\mscript{c} ({\rm kpc}) < 10^3$. 

Figure \ref{fig:plotNFWburkert} shows the distribution of $r_\mscript{cn}$ values as found from the individual fits.  Again, as expected, the NAV analysis can be used to infer an approximate range of $r_\mscript{cn}$ values that are most common within the individual fits (this independently on whether the stellar mass-to-light ratios are being fitted or are fixed at their central values).

The median of the minimum $\chi^2$ value for the Burkert halo reads 
$\widetilde{\chi^2_\mscript{Bur}} \approx 8 $. This is the lowest median $\chi^2$ of the models considered up to here. As expected from the NAV analysis, this $\chi^2$ result is much better than that of the NFW halo or the Arctan$_{1/2}$ model.

\subsection{Modified Newtonian Dynamics (MOND)} \label{sec:ModelMOND}

\subsubsection{Model definition} \label{sec:ModelDefMOND}

MOND is a well-known and simple modified gravity model that, contrary to several other proposals, has achieved certain success in the context of galaxy rotation curves \citep{1983ApJ...270..371M, Famaey:2011kh}. Following the original MOND formulation, as well the more recent motivation from the Radial Acceleration Relation (RAR) \citep{McGaugh:2016leg}, this model imposes that the observational acceleration can be expressed as a function of the Newtonian acceleration,
\begin{equation} \label{eq:mondMu}
	\mathbf a_\mscript{N} = \mu\left (\frac{a}{a_0} \right ) \mathbf a \, ,
\end{equation}
where $a_0$ is a constant, $\mu$ is called the interpolation function, $\mathbf a_\mscript{N}$ is the Newtonian acceleration inferred from baryonic matter and $\mathbf a$ is the observational one. Within axis-symmetric disk galaxies, the acceleration vectors $\mathbf a$ and $\mathbf a_\mscript{N}$ are along the radial direction in the plane of the disk ($z=0$), hence $\mathbf a = - \frac{V^2}{r} \hat{\boldsymbol r}$ and  $a = |\mathbf a| = \frac{V^2}{r}$.

The interpolation function $\mu$ must satisfy $\mu(x) = 1$ for $x \gg 1$ and $\mu(x) = x$ for $x \ll 1$. For $x \sim 1$, the interpolation function behaviour should be fixed from the observational data. The RAR suggests the following relation between $\mathbf a$ and $\mathbf a_\mscript{N}$ \citep{McGaugh:2016leg},
\begin{equation} \label{eq:mondRAR}
	\mathbf a = \frac{\mathbf a_\mscript{N}}{1 - e^{- \sqrt{a_{\rm N}/a_0}}} \, .
\end{equation}
The same data lead to $a_0 = 1.2 \times 10^{-10}$ m/s$^2$ as the best fit for $a_0$ (considering all the galaxies together). Another possible value for $a_0$, from the same galaxies and the same function \eqref{eq:mondRAR} was derived by \cite{Rodrigues:2018duc, Marra:2020sts}, where the best value for $a_0$ is selected as the one that minimizes the tensions between the individual galaxies, considering the observational uncertainties, leading to lower value, $a_0=9.6 \times 10^{-11}$ m/s$^2$, but still in accordance with $a_0 \sim 10^{-10}$ m/s$^2$.

MOND is here considered since it is well-known, simple and since it provides interesting results for the NAV analysis. Some of us have shown that  the observational uncertainties are not compatible with the assumption that there is a common $a_0$ value for all the galaxies, \citep{Rodrigues:2018duc, Marra:2020sts}.  The issue was found with a high significance, always higher than $5 \sigma$. These results were further discussed by \citet{McGaugh:2018aa, Kroupa:2018kgv, Rodrigues:2018lvw, 2020NatAs...4..132C, Rodrigues:2020squ, Rodrigues:2020gbg, Li:2021ted}, see also \cite{Chang:2018lab, 2019ApJ...882....6S,  Ren:2018jpt, Zhou:2020tst, 2021A&A...656A.123E}. This is a problem for MOND [the original MOND version, which is based on eq.~\eqref{eq:mondMu}] as a correct theory for galaxies \citep[see also][]{Navarro:2016bfs, 2019ApJ...882....6S, Dutton:2019gor}. 

The NAV analysis, as here presented, is not concerned with the basis of MOND, but it stresses a peculiar trend in MOND's rotation curves: its lack of diversity \cite[see also][for similar conclusions from different methods]{Ren:2018jpt, Kaplinghat:2019dhn}.

\subsubsection{NAV analysis} \label{sec:NAVmond}

From eq.~\eqref{eq:mondRAR} with $a_\mscript{N}= a_\mscript{bar} = \Vbar / r$, the $\Delta \Vmod$ MOND expression reads
\begin{equation} \label{DeltaV2MOND}
	\Delta V^2_\mscript{MOND} =  V^2_\mscript{MOND} - \Vbar = \frac{ \Vbar }{ e^{\sqrt{a_\mscript{bar} / a_0} } - 1} \, ,
\end{equation}
hence
\begin{equation} \label{deltaV2Mond}
	\delta V^2_\mscript{MOND}(\rn) = \frac{\Vbar(\rn)}{\Vbar(1)} \frac{e^{\sqrt{a_\mscript{bar}(1) / a_0}}-1}{e^{\sqrt{a_\mscript{bar}(\rn) / a_0}}-1} \, .
\end{equation}

Since MOND is nonlinear on the constant $a_0$, this constant is still present in $\delta \Vmod$. 

The Newtonian limit of MOND is found in the limit $a_0 \to 0$. Assuming that $\Vbar(\rn) > 0$, then, for a given $\rn \in  (0,1)$, 
\begin{equation}
	\lim_{a_0 \to 0} \delta V^2_\mscript{MOND} (\rn) = 
	\begin{cases}
	  0 \, , & \mbox{if } a_\mscript{bar}(1) < a_\mscript{bar}(\rn)\\[.1cm]
	  \infty	 \, , & \mbox{if } a_\mscript{bar}(1) > a_\mscript{bar}(\rn)
	\end{cases}	\, .
\end{equation}

Hence, the Newtonian limit of MOND leads to $\delta V^2_\mscript{MOND} = 0$ for either all or most of the $\rn$ values. We recall that, in general, $\Vmod$ is ill defined for Newtonian gravity without dark matter. The limit of a modified gravity theory towards Newtonian gravity can display different features depending on the considered theory.

On the other hand, the limit $a_0 \to \infty$ implies (assuming $\Vbar > 0$)
\begin{equation} \label{MONDsqrt}
	\lim_{a_0 \to \infty} \delta V^2_\mscript{MOND}(\rn) =  \sqrt \rn \frac{V_\mscript{bar}(\rn)}{V_\mscript{bar}(1)} \, .
\end{equation}
This type of regime of large $a_0$ (or small $a_\mscript{N}$) is commonly called the ``deep MOND regime''. For large radii, most of the galaxies are expected to be in this regime, also $V_\mscript{bar}(1) \sim V_\mscript{bar}(1)$, and hence $\delta V^2_\mscript{MOND}(\rn) \sim \sqrt \rn$. It should be noted that this behaviour is close to the best-fit curve found from the polynomial approach \eqref{eq:polyModels}. Hence, MOND should be good to cover the central region of the NAV plane. On the other hand, there is no independent mechanism to adjust the data distribution, the 2$\sigma$ region in particular. Indeed, as it will be verified numerically, MOND will display a lack of diversity.

\begin{figure*}
	\begin{tikzpicture}
  		\node (img1) {
  			\includegraphics[height=0.27\textwidth, trim = 0 0 0 0, clip]{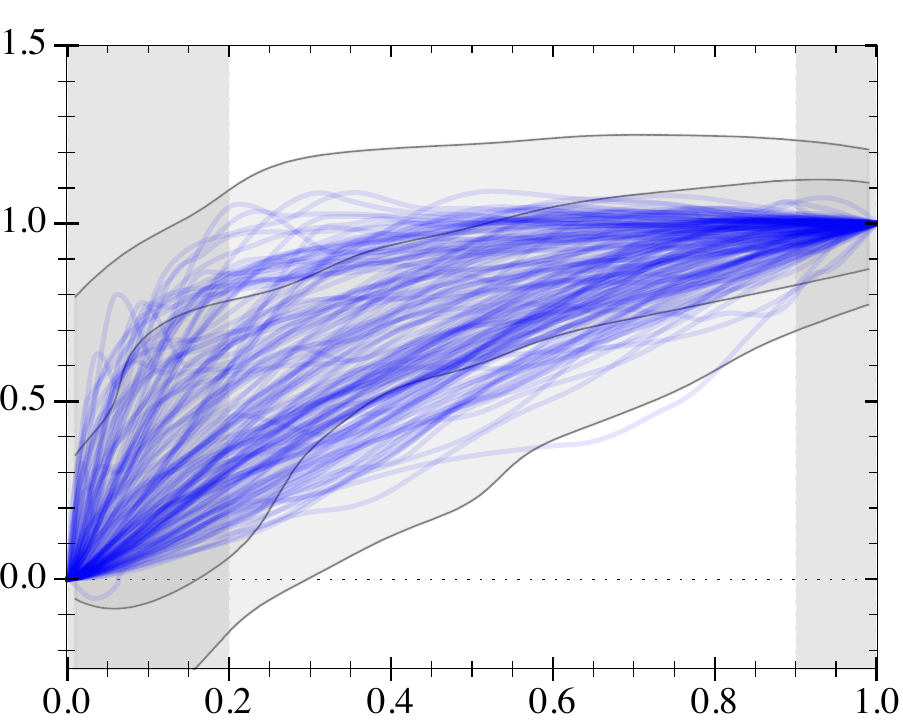}
  		};
		\node[below=of img1, node distance=0cm, yshift=1.1cm, xshift=0.15cm] {
			\large $\rn$
		};
		\node[left=of img1, node distance=0cm, rotate=90, yshift=-0.9cm, xshift=0.6cm] {
			$\delta V^2$
		};
		\node[above=of img1, node distance=0cm, yshift=-1.4cm, xshift=0.15cm] {
			$a_0 = 1 $ km/s$^2$
		};
	\end{tikzpicture} \hspace{-0.51cm}
	\begin{tikzpicture}
  		\node (img2) {
  			\includegraphics[height=0.27\textwidth, trim = 19 0 0 0, clip]{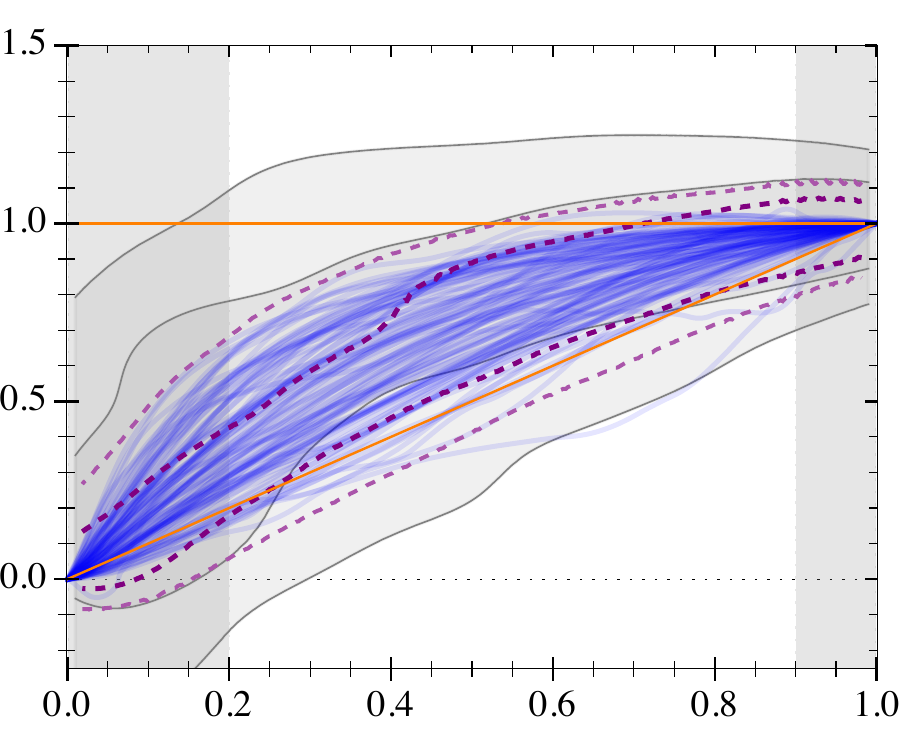}
  		};
		\node[below=of img2, node distance=0cm, yshift=1.1cm, xshift=0.0cm] {
			\large $\rn$
		};
		\node[above=of img2, node distance=0cm, yshift=-1.4cm, xshift=0.0cm] {
			$a_0 = 1.2 \times 10^{-13}$ km/s$^2$
		};
	\end{tikzpicture}\hspace{-0.44cm}
	\begin{tikzpicture}
  		\node (img3)  {
  			\includegraphics[height=0.27\textwidth, trim = 19 0 0 0, clip]{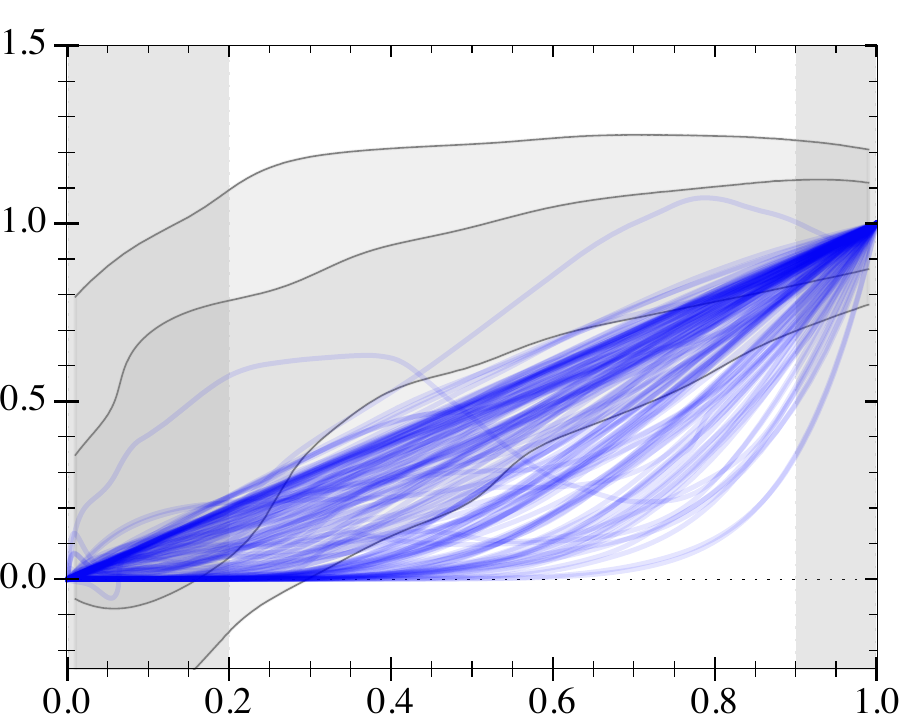}
  		};
		\node[below=of img3, node distance=0cm, yshift=1.1cm, xshift=0.0cm] {
			\large $\rn$
		};
		\node[above=of img3, node distance=0cm, yshift=-1.4cm, xshift=0.0cm] {
			$a_0 = 10^{-15}$ km/s$^2$
		};
	\end{tikzpicture}
	\caption{The NAV  for MOND, as given by eq.~\eqref{eq:mondRAR}, for three different $a_0$ values. Each transparent blue curve corresponds to one of the 153 SPARC galaxies, the same ones that were used for the RAR \citep{McGaugh:2016leg}. The central plot assumes the commonly used $a_0$ value ($1.2 \times 10^{-13}$ km/s$^2$), while the other two plots assume extrema cases (1 km/s$^2$ and $10^{-15}$ km/s$^2$). The central plot includes the limits of the 1$\sigma$ and 2$\sigma$ HDRs, which are shown in dashed purple curves. It also includes, for reference, the orange straight lines $\delta V^2 = \rn$ and $\delta V^2 = 1$, which are the limits of the arctan model for $\alpha = 1/2$. The mean MOND efficiency for the central plot is 0.53 \eqref{meanEfficiencyMOND}.}
	\label{fig:MONDraw} 
\end{figure*}
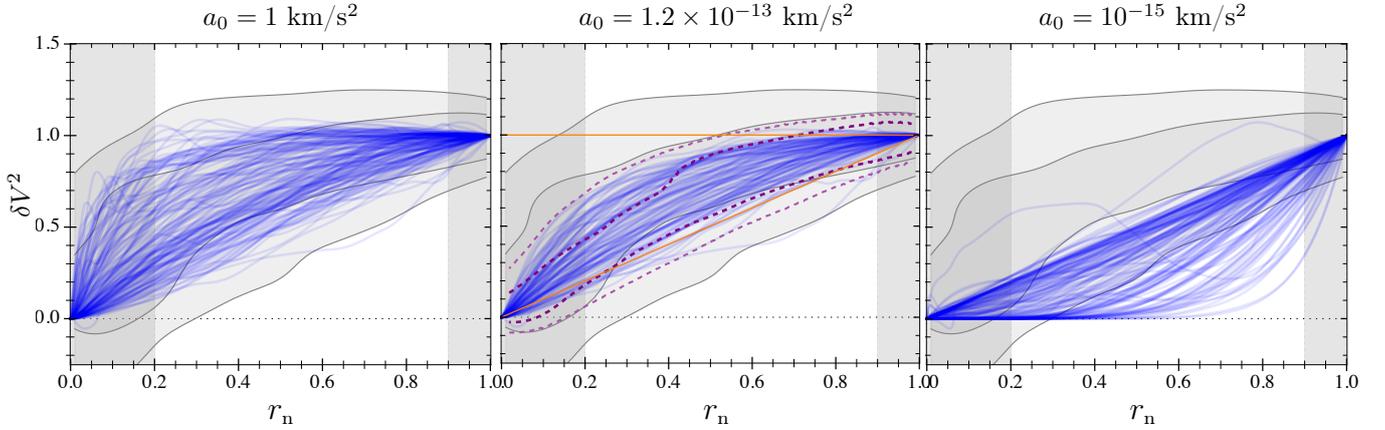

Contrary to the previous dark matter cases, the $\delta V^2_\mscript{MOND}(\rn)$ curves directly depend on $\Vbar(\rn)$. Hence, to consider the 1$\sigma$ and $2\sigma$ HDR's for MOND, we use eq.~\eqref{deltaV2Mond} to draw the $\delta V^2_\mscript{MOND}(\rn)$ curves for each one of the 153 $\Vbar(\rn)$ curves. These data are provided by SPARC and the resulting curves are plotted in Fig.~\ref{fig:MONDraw}. After that, these curves are randomly discretized, to smooth out the data points distribution, and the 1$\sigma$ and 2$\sigma$ HDR's are computed. These regions are also plotted in the central plot of Fig.~\ref{fig:MONDraw}. Besides this plot, we also show the consequences of considering either very large or very small $a_0$ values. 

The NAV efficiency for the MOND model reads
\begin{equation}
  E_1 = 0.515 \, , \;  E_2 = 0.544 \, \mbox{, and }  E_\mscript{M} = 0.53\,  . \label{meanEfficiencyMOND}
\end{equation}
This is the lowest efficiency value here computed. Considering Fig.~\ref{fig:MONDraw}, it is not a surprising result, since the MONDian curves are almost entirely inside the region delimited by the arctan model with $\alpha=1/2$. The 1$\sigma$ and 2$\sigma$ regions from MOND are narrower than the observational ones.

As the first plot of Fig.~\ref{fig:MONDraw} shows, considering the NAV plane covering alone, a large value for $a_0$ is favoured. Such large $a_0$ value in on the other hand incompatible with the $\Delta V^2_\mscript{obs}$ data, Fig.~\ref{fig:DeltaV2Analysis}.

\subsubsection{Comparison with the individual fits results} \label{sec:MONDindividual}

For MOND, similarly with the previous sections, we fitted the galaxies individually. The single parameters to be fitted, in this case, were the disk and bulge stellar mass-to-light ratios. We assumed the usual $a_0$ value, that is $a_0 = 1.2 \times 10^{-13}$ km/s$^2$. This model low $E_\mscript{M}$ \eqref{meanEfficiencyMOND} suggests that MOND will have a large median $\chi^2$ value. And, indeed,    $\widetilde{\chi^2_\mscript{MOND}} \approx 48$.

Among the  models here considered, this is the one with largest $\widetilde{\chi^2}$ value, which is expected since it also has the smallest $E_\mscript{M}$ value. It should be pointed out that this is the model with least free parameters, it only depends on the global $a_0$ value. We are not using any metric for compensating for the number of parameters, it is not our purpose here.

MOND has no additional parameter that was not considered in the NAV analysis, this is different from the other models. For this model individual fits, we have also done  fits with distance variations compatible with the uncertainties reported by SPARC. Within this case, the median value is reduced to $ \widetilde{\chi^2_\mscript{MOND}} \approx 30$.

\subsection{DC14} \label{sec:DC14}
\subsubsection{Model definition} \label{sec:ModelDefDC14}

The \citet{DiCintio:2014xia} model (DC14) extends the NFW halo by considering baryonic feedback.  The DC14 halo profile depends on the stellar mass of the galaxy ($M_*$), more specifically, on the quantity $X = \log_{10} M_* / M_\mscript{halo}$, where $M_\mscript{halo}$ is the dark matter halo mass.

\begin{align}
  &\rho_{\mscript{DC14}}(r) =  \frac{\rho_\mscript{s}}{\frac{r^\gamma}{r_s^\gamma} \left (  1 + \frac{r^\alpha}{r_\mscript{s}^\alpha}  \right )^{(\beta-\gamma)/\alpha }} \, , \nonumber\\
  &\alpha =  2.94\, -\log _{10}\left(\left(10^{X+2.33}\right)^{2.29}+\frac{1} {\left(10^{X+2.33}\right)^{1.08}}\right) \, , \nonumber\\
  &\beta =  0.26 X^2+1.34 X+4.23 \, , \label{dc14def}\\
  &\gamma =  \log _{10}\left(10^{X+2.56}+\frac{1}{\left(10^{X+2.56}\right)^{0.68}}\right)-0.06 \, . \nonumber
\end{align}
The parameters above were derived for $-4.1 < X < -1.3$. The case $X = -4.1$ leads to the NFW halo (within a good approximation).  

From its definition above, each $\delta V^2_\mscript{DC14}$ curve depends on $X$ and $r_\mscript{s}$. $X$ can be estimated  for each galaxy from the stellar to halo mass relation \citep{Moster:2012fv}. We do not provide here individual fits for this model, but they were considered, using the SPARC data, by \citet{Katz:2016hyb, Li:2020iib}. Besides other improvements, it was found that the overall quality of the fits is significantly better than the NFW profile and quite similar to the Burkert profile results (considering their CDF plots). Here we explore what kind of improvements DC14 has over NFW in the NAV plane, and to which extent it is close to the Burkert profile.

\subsubsection{NAV analysis} \label{sec:NAVDC14}

From eq.~\eqref{dc14def}, one can define $\delta V^2_\mscript{DC14}(\rn, r_\mscript{s}, X)$. As it was shown in Sec.~\ref{sec:NFW}, two relevant problems that the NFW halo has are: $i$) the lower limit in the NAV plane of the NFW halo is given by the curve $\delta V^2_\mscript{NFW} = \rn$, which is too high. $ii$) To achieve this lower limit it is necessary to use very large $r_\mscript{s}$ values (consequently,  very low values for the concentration $c$ are found from real galaxies data, being at odds with the simulation results). 

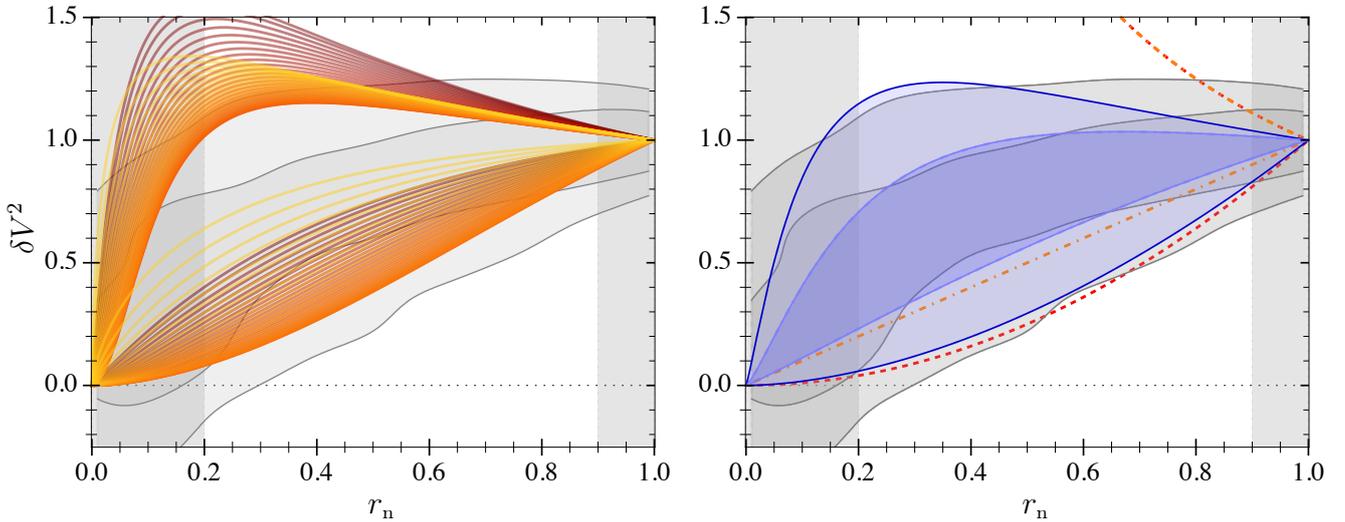
\begin{figure*}
	\begin{tikzpicture}
        \node (img1)  {\includegraphics[width=0.45\textwidth]{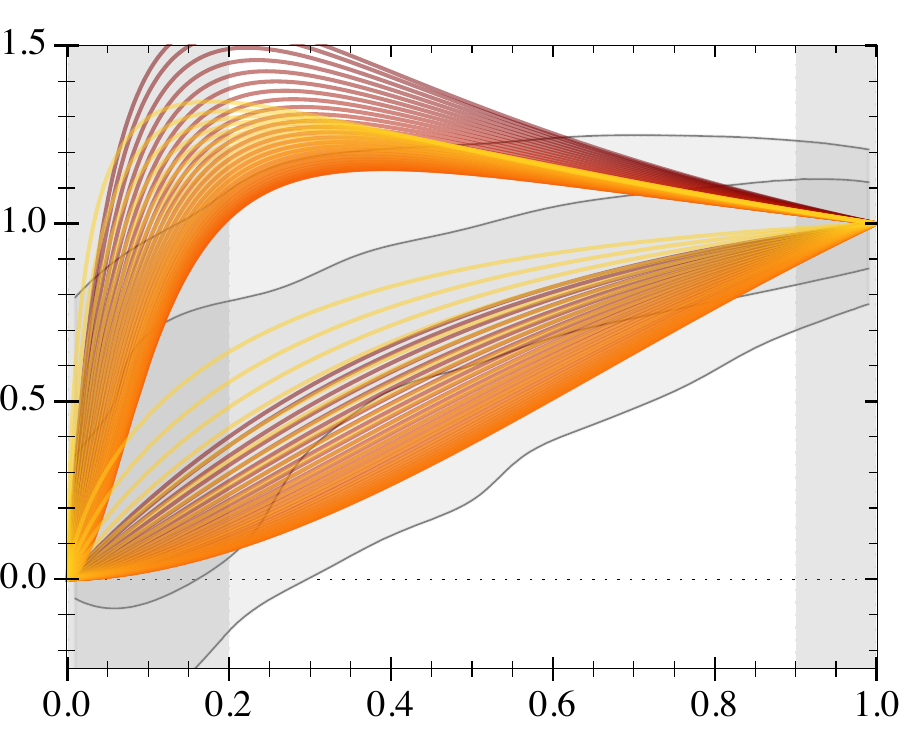}};
		\node[below=of img1, node distance=0cm, yshift=1.1cm, xshift=0.3cm, font=\color{black}] {\large $\rn$};
		\node[left=of img1, node distance=0cm, rotate=90, yshift=-0.8cm, xshift=0.6cm] {\large $\delta V^2$};
	\end{tikzpicture}
	\begin{tikzpicture}
  		\node (img1)  {\includegraphics[width=0.45\textwidth]{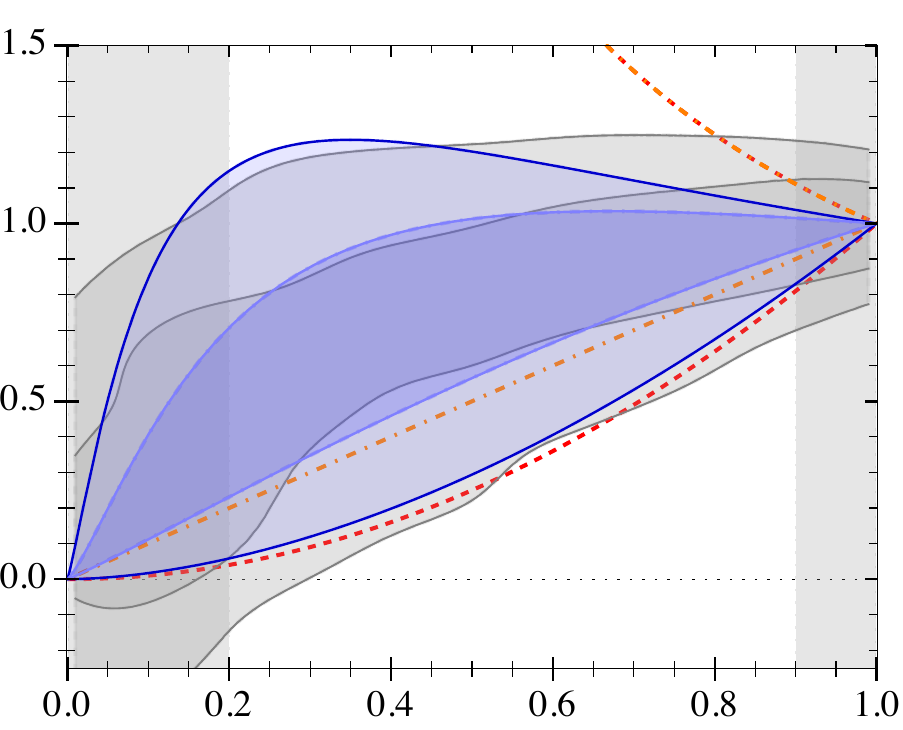}};
		\node[below=of img1, node distance=0cm, yshift=1.1cm, xshift=0.3cm, font=\color{black}] {\large $\rn$};
	\end{tikzpicture}
	\caption{
      \textbf{Left.} The curves $\delta V^2_\mscript{DC14}$ for two values of $r_\mscript{sn}$: 0.1 and 1, and for several values of $X$, with different colors. The curves in the upper part of the plot use $r_\mscript{sn} = 0.1$, while those in the lower part use $r_\mscript{sn} = 1$. From dark red to light yellow, $X$ changes from $- 4.1$ to $-1.3$.
      \textbf{Right.} The NAV plane for the DC14 model. The orange dot-dashed curve shows the  the NFW profile extrema cases \eqref{nfwExtrema}, while the red dashed curve shows the same for the Burkert profile \eqref{burkertExtrema}. The two bluish regions are the DC14 model best approximation for the observational 1$\sigma$ and 2$\sigma$ regions \eqref{rsnDC14HDRLimits}. The mean efficiency is 0.80 \eqref{meanEfficiencyDC14}.
	}
	\label{fig:plotDC14} 
\end{figure*}

The two issues above can be directly shown to be absent from the DC14 halo using the NAV analysis. In Fig.~\ref{fig:plotDC14} we show $\delta V^2_\mscript{DC14}$ curves for two $r_\mscript{sn}$ values ($r_\mscript{sn}= 0.1$ and $r_\mscript{sn} = 1$) and for different possible $X$ values. For the NFW halo, $r_\mscript{sn}=0.1$ corresponds to the darkest curve in the lower part of Fig.~\ref{fig:plotDC14}, hence far from the 2$\sigma$ limit and still inside the 1$\sigma$ region. For the DC14 model, by changing the value of $X$, it is possible to go closer to the 2$\sigma$ limit than any NFW curve can do (even in the limit $r_\mscript{sn} \to \infty$). This observation does not show that DC14 will necessarily work for individual galaxies,  but it is sufficient for showing that DC14 can overcome the problems that NFW halo has for any galaxy. 

It is also possible to fit $\delta V^2_\mscript{DC14}$ to the 1$\sigma$ and 2$\sigma$ HDRs, as follows:

\begin{align}
	&2\sigma_+:  r_\mscript{sn} = 0.14 \, , \; X = -3.88 \, , \nonumber\\ 
	&1\sigma_+: r_\mscript{sn} = 0.24 \, , \; X = -3.76 \, ,\nonumber\\ 
	&1\sigma_-: r_\mscript{sn}= 1.91 \, , \; X = -1.60 \, ,\label{rsnDC14HDRLimits} \\  
	&2\sigma_-: r_\mscript{sn} = \infty \, , \; X = -2.66 \, .\nonumber 
\end{align}
The corresponding curves are plotted in Fig.~\ref{fig:plotDC14}.

\begin{figure}
	\begin{tikzpicture}
  		\node (img1)  {\includegraphics[width=0.45\textwidth]{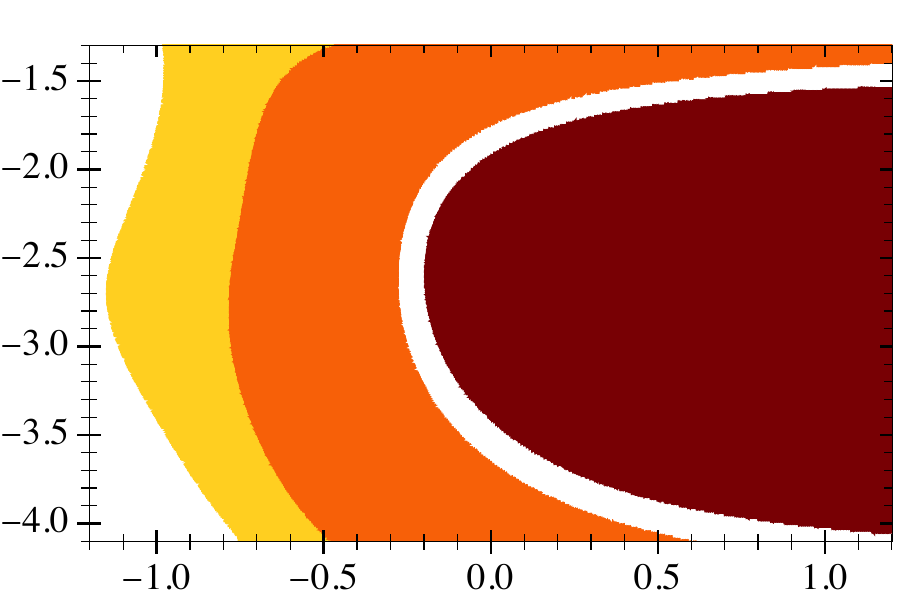}};
		\node[below=of img1, node distance=0cm, yshift=1.1cm, xshift=0.3cm, font=\color{black}] {\large $\log_{10} r_\mscript{sn}$};
		\node[left=of img1, node distance=0cm, rotate=90, yshift=-0.8cm, xshift=0.6cm] {\large $X$};
	\end{tikzpicture}
	\caption{Five regions of the DC14 parameter space: the leftmost white region  corresponds to the parameter region in which the curves $\delta V^2_\mscript{DC14}$ are above the $2\sigma_+$ curve \eqref{rsnDC14HDRLimits}. The yellow region corresponds to $\delta V^2_\mscript{DC14}$ curves above  $1\sigma_+$ but below $2\sigma_+$. The next region,  in orange, corresponds to the $\delta V^2_\mscript{DC14}$ curves inside the 1$\sigma$ region. The thin white region is the region in which $\delta V^2_\mscript{DC14}$ is between $1\sigma_-$ and the asymptotic lowest limit of $\delta V^2_\mscript{NFW}$. The rightmost region is in dark red and it corresponds to the $\delta V^2_\mscript{DC14}$ curves that are below the lowest possible $\delta V^2_\mscript{NFW}$ curves. All comparisons are done at $\rn = 0.5$.}
	\label{fig:plotRegionsDC14} 
\end{figure}

For fixed $r_\mscript{sn}$ and $\rn$, the curves $\delta V^2_\mscript{DC14}(X)$ are not monotonous, hence it is not possible to directly use the fits above to state parameter regions.  Figure~\ref{fig:plotRegionsDC14} shows the parameter space regions that are either above 2$\sigma_+$, between $2\sigma_+$ and $1\sigma_+$, inside the 1$\sigma$ region, between $1\sigma_-$ and the lowest possible $\delta V_\mscript{NFW}$ curve, and those curves that can go below the  $\delta V_\mscript{NFW}$ ones. The plot was done by classifying the curves at the fixed radius $\rn = 0.5$. 

Figure \ref{fig:plotRegionsDC14} shows that any stellar mass (within the considered range by \citet{DiCintio:2014xia}) can be used to cover either the 1$\sigma$ region or the region above it. For the region below the 1$\sigma$ one, it is possible to have reasonable $r_\mscript{s}$ values if $X$ is in the range $-2.0 \lesssim X \lesssim -3.5$. Also, one concludes that it is possible to go beyond the limitations of NFW in the NAV plane with reasonable $r_\mscript{s}$ values. We recall that NFW corresponds approximately to the case $X = -4.1$.

Using the curves from eq.~\eqref{rsnDC14HDRLimits} one can compute the NAV efficiency of the DC14 model as
\begin{equation}
  E_1 = 0.776 \,, \;  E_2 = 0.824 \, \mbox{, hence } \; E_\mscript{M} = 0.80\, . \label{meanEfficiencyDC14}
\end{equation}
This is the same $E_\mscript{M}$ that was found for Burkert. 

We do not compute the individual DC14 fits here, but considering the $\delta V^2_\mscript{DC14}$ curves and the efficiency values above, the median $\chi^2$ values should be about the same of those from Burkert and the arctan model with $\alpha = 1$. Indeed, \citet{Li:2020iib} show that the DC14 fits are on average clearly better than those of NFW and very close to the Burkert profile fits.

\section{Model comparisons} \label{sec:modelComparisons}

In this section we summarize and compare the models results.

Table \ref{tab:NAVefficiency} shows the mean efficiency ($E_\mscript{M}$) results for each model, together with the corresponding medians of $\chi^2$ and  $\chi^2_\mscript{eff}$. The medians are denoted by a tilde and the definitions of the previous quantities are in Sec.~\ref{sec:individual}.  One sees that there is a correlation between $E_\mscript{M}$ and the median values of $\chi^2$ and $\chi^2_\mscript{eff}$.

Among the models with individual galaxy fits, there is no surprise that the Burkert profile has the best score for both $E_\mscript{M}$ and $\chi^2$ in Table \ref{tab:NAVefficiency}. It is a model known to be capable of providing good fits to a large number of galaxies, usually being better (i.e., with lower $\chi^2$ values) than the NFW profile and modified gravity approaches \citep[e.g.,][]{Gentile:2004tb, Rodrigues:2014xka, Li:2020iib}.

As detailed in Sec.~\ref{sec:individual}, we recall that the NAV analysis considers $\YD$ and $\YB$ as fixed, but the sample results are also valid for $\YD$ and $\YB$ with Gaussian priors. This is expected since  there is no reason for a correlation between variations of the stellar mass-to-light ratios and regions in the NAV plane, nor reasonable models are expected to introduce a bias towards particular $\YD$ values. We have also directly verified that, for all the models here tested, the $\logYD = \log_{10} \YD$ distribution from the fits is compatible with Gaussian and centered at  $\sim \log_{10} 0.5$. 

For MOND, since it is a model particularly susceptible to small changes in the baryonic content, and since this model does not have an independent parameter that controls the magnitude of the non-Newtonian effects, we have also considered changes in the galaxy distances that are compatible with the observational errors. When such changes are considered, we labeled the model as MOND$_\mscript{dist.}$. This additional freedom  reduces the median $\chi^2$ by several units, but the final $\widetilde \chi^2$ result is still larger than the Arctan$_{\alpha=1/2}$ model. Details on the distance modeling can be found for instance in the supplementary material of \citet{Rodrigues:2018duc} or \citet{Li:2018tdo, Marra:2020sts}. One could consider galaxy inclination changes as well, and the median $\chi^2$ for this dataset would be somewhat lower, but without strong changes for the sample results (since changing $\Upsilon_*$ and the distance are more relevant for this sample \citep{McGaugh:2016leg, Rodrigues:2018lvw}). Either with or without distance variations, we are finding a correlation between the efficiency and $\chi^2$.

\setlength{\tabcolsep}{10pt}
\begin{table}
\begin{center}
\caption{Model mean NAV efficiency ($E_\mscript{M}$), median $\chi^2$ and $\chi^2_\mscript{eff}$ results. The models are sorted from larger to smaller $E_\mscript{M}$. Larger $E_\mscript{M}$ values indicate better agreement between theory and observation in the NAV plane (Sec.~\ref{sec:ModelEfficiencyIntro}). \\}
\label{tab:NAVefficiency}
\begin{tabular}{lrrr}
\hline 
\hline \\[-.3cm]
Model                 &$E_\mscript{M}$ & $\widetilde{\chi^2}$ & $\widetilde{\chi^2_\mscript{eff}}$\\[0.1cm]
\hline
Polynomial            & 0.86 & ---  & ---\\
DC14                  & 0.80 & ---  & ---\\
Burkert               & 0.80 & 8.0  & 8.1\\
Arctan$_1$            & 0.70 & 8.5  & 8.9\\
NFW                   & 0.70 & 14.9 & 16.0\\
Arctan$_{1/2}$        & 0.60 & 17.7 & 19.0\\
MOND$_\mscript{dist.}$& 0.53 & 29.6 & 37.1\\
MOND                  & 0.53 & 47.6 & 59.6\\
\hline 
\hline
\end{tabular}
\end{center}
\end{table}

\begin{figure}
	\begin{tikzpicture}
  		\node (img1)  {\includegraphics[width=0.45\textwidth]{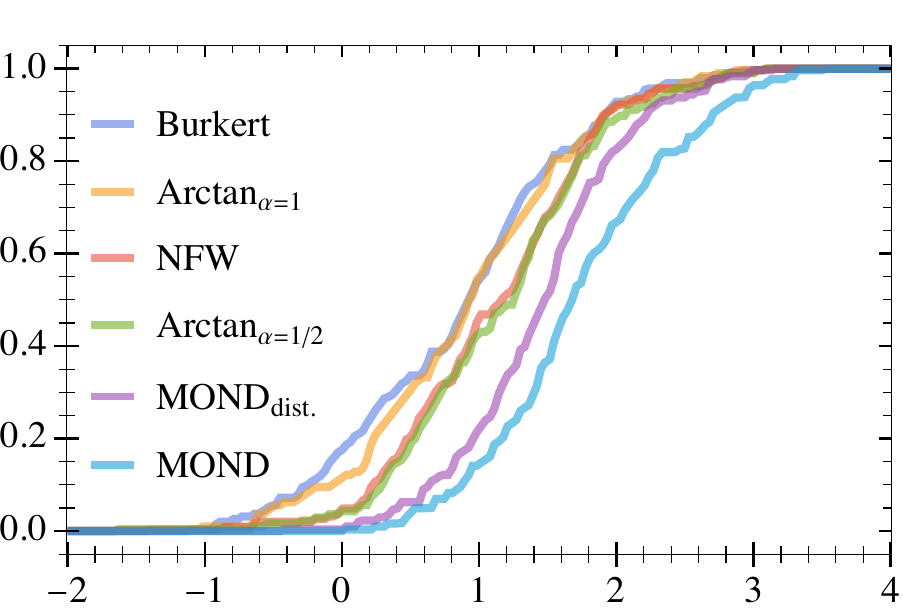}};
		\node[below=of img1, node distance=0cm, yshift=1.1cm, xshift=0.3cm, font=\color{black}] {\large $\log_{10}\chi^2$};
		\node[left=of img1, node distance=0cm, rotate=90, yshift=-0.8cm, xshift=0.6cm] {\large CDF};
	\end{tikzpicture}
	\caption{The CDFs for the $\chi^2$-values found from the individual fits of different models.  The median ${\chi^2}$ values, as displayed in Table~\ref{tab:NAVefficiency}, can be found at the intersection between the curves and the CDF value of $0.5$. This plot generalizes the previous table results, since it yields results for any percentile, not only the median.}
	\label{fig:plotCDFcomparison} 
\end{figure}

Table \ref{tab:NAVefficiency} suggests that the larger is $E_\mscript{M}$ the lower is the median $\chi^2$. Figure~\ref{fig:plotCDFcomparison} shows the CDF plots for $\chi^2$, considering all the models that were here fitted with individual galaxies. The $\widetilde{\chi^2}$ results in Table~\ref{tab:NAVefficiency} correspond to the horizontal line for which the CDF is equal to 0.5. Hence the figure shows that the $\widetilde{\chi^2}$ ordering in Table~\ref{tab:NAVefficiency} is not an accident related to the use of the median, the same order could be found at almost any CDF value (or, equivalently, at any adopted percentile, not only the median). The CDF plots for $\chi^2_\mscript{eff}$ are qualitatively the same.

\section{Conclusions and discussions} \label{sec:conclusions}

We developed a method, based on the normalized additional velocity (NAV) \eqref{NAVdef}, to study galaxy RCs directly from a sample of galaxies. This approach is inspired by an analysis of \citet{Rodrigues:2009vf, Rodrigues:2011cq}, and in part it is a normalized version of the approach of \citet{McGaugh:2006vv}. With respect to the latter, we focus on the radial dependence of the expected dark matter contribution, or, in the case of modified gravity, in  the non-Newtonian  contribution. This is done by eliminating the amplitude dependence through a normalization. The method was not developed to be a complete data assessment, but to provide an efficient sample comparison of a particular and important aspect of galaxy rotation curves (RCs): their shape. Among its applications that are here illustrated, this method can be used to eliminate model parameter regions, to find the most probable parameter regions, to uncover trends that are not simple to be found from standard individual fits and to compare models.  

Considering the SPARC galaxy sample \citep{2016AJ....152..157L}, the essential observational data to be used for the NAV evaluation is provided in tabular data in Table~\ref{tab:tabularsigmas}, see also Fig.~\ref{fig:plotBlueRAR}. For some models, the method can be relevant even analytically. Also, the fact that this approach commonly eliminates one model parameter should not be underestimated, since this simplifies the analysis and allows for data-based inferences on the model parameters that are independent from the eliminated one.

After explaining that there should be a correlation between the NAV analysis and the average minimum $\chi^2$ values, as discussed in Sec.~\ref{sec:method}, we have confirmed  such a correlation in Table~\ref{tab:NAVefficiency}.  

Five models were here considered, namely: three well-known dark matter profiles (NFW, Burkert and the DC14 profiles), a well-known modified gravity model (MOND) and a more phenomenological model here called Arctan$_\alpha$. The Arctan$_\alpha$ model coincides with the arctan model studied by \citet{Courteau:1997wu} for the special case $\alpha = 1$. The main purpose of this Arctan$_\alpha$ model was to test the NAV method, since this model for $\alpha = 1$ and $\alpha = 1/2$ is similar to the Burkert and NFW models respectively, considering the lower part of the NAV plane (as it can be seen in Figs.~\ref{fig:plotArctan}, \ref{fig:plotNFWburkert}).

We have shown that the simple Arctan$_{\alpha = 1}$ model is indeed quite efficient on providing good rotation curve fits. We have not explicitly evaluated the pseudo-isothermal model here \citep[e.g.,][]{1991MNRAS.249..523B}, but its results are expected to be similar. As anticipated from the NAV analysis and confirmed from the individual fits, the Arctan$_{\alpha = 1/2}$ is significantly worse than the latter. 

The method was also sufficient for showing that, for the SPARC galaxies, the Burkert halo performs on average significantly better than NFW and Arctan$_{\alpha = 1/2}$. The NAV result is not related with the core-cusp issue in its stricter sense (i.e., at regions very close to the galaxy center), but to the necessary diversity of dark matter halos that NFW lacks \citep{Oman:2015xda}. The Burkert profile can reproduce such diversity, but without a proper physical context. The physical picture is improved with the DC14 halo, and the NAV analysis also show that this model is capable of providing a diversity of halo shapes similar to the Burkert profile, also in agreement with \citet{Katz:2016hyb, Li:2020iib}. In Fig.~\ref{fig:plotRegionsDC14} we show DC14 parameter regions that are more common, and those values that are needed to go beyond the NFW problems, considering the SPARC galaxies.

We have also pointed out that MOND in its original form can cover well the central region of the NAV plane [see Fig.~\ref{fig:MONDraw} and also eq. \eqref{MONDsqrt}], but it has an issue qualitatively similar to the NFW halo: it lacks diversity, hence the individual fits lead to large $\chi^2$ values. These results of ours are similar with the results from \citet{Ren:2018jpt, 2021A&A...656A.123E}, although the conclusions were found from a  different method.

The method here presented can be extended to be applied in more general modified gravity models. The application to MOND (in its original form) is straightforward since, contrary to other modified gravity models, its rotation curve can be written as a function of $\Vbar$. For other modified gravity models, commonly one needs to know the three dimensional baryonic matter distribution \citep[see for instance][for further details]{Green:2019cqm}. We understand that this method can bypass certain difficulties on the analysis of modified gravity. We plan to extend it in this direction as well.

\section*{Acknowledgments} 
\noindent
DCR thanks CNPq and FAPES for partial financial support. AHA acknowledges financial support from CAPES and DAAD. AW acknowledges financial support from the EU through the European Regional Development Fund CoE program TK133 ``The Dark Side of the Universe" and from from MICINN (Spain) {\it Ayuda Juan de la Cierva - incorporac\'on} 2020 No. IJC2020-044751-I.

\section*{Data Availability}
\noindent
The data underlying this article are available either in its online supplementary material, in the \texttt{NAVanalysis} code \citep{NAVanalysis}, or in \href{http://astroweb.cwru.edu/SPARC/}{http://astroweb.cwru.edu/SPARC/} \citep{2016AJ....152..157L}.

\section*{Author contributions}
\noindent
DCR: conceptualization, methodology, software, writing. AHA: software, methodology. AW: conceptualization, writing. 
 

\appendix

\section{KDE bandwidth selection} \label{app:KDE}

In order to find the distributions used in this work, it was necessary to convert data points into distributions. A well established procedure is to use a kernel density estimation (KDE). To use it, one needs to specify a kernel and a bandwidth.  It is well known that a Gaussian kernel is both convenient and it has a high efficiency, with respect to the mean integrated square error (MISE) \citep{Silverman_1998, Scott_2014}. The Gaussian kernel is always assumed here. 

Although the bandwidth for several problems can be chosen by eye, for this work it is important to use a single algorithm to select it, thus avoiding any human bias on the bandwidth selection. Here we opted for the Silverman rule \citep{Silverman_1998}, which is a simple and standard bandwidth selection rule. For consistency, every time a KDE is used here, we use the Gaussian kernel with the Silverman rule.

We start by considering the unidimensional problem. This brief review is mainly based on \citet{Silverman_1998}. Let $\{x_i\}$ be a set of $n$ observations whose probability density is denoted by $f(x)$ and it is \textit{a priori} unknown.  A KDE estimator $\hat f$ for the distribution $f$ is given by
\begin{equation}
	\hat f (x) = \frac 1 {n h} \sum_{i=1}^n K\left ( \frac{x - x_i}{h}\right) \, ,
\end{equation}
where $h$ is the bandwidth and $K$ is the kernel. The latter satisfies
\begin{equation}
	\int_{-\infty}^\infty K(x) dx = 1 \, .
\end{equation}
This condition ensures the normalization of $\hat f$. Therefore, a KDE with a Gaussian kernel is a sum of normalized Gaussians, one Gaussian for each data point.

The bandwidth ($h$) selection is the trickiest part. This since the variation of $\hat f$ decreases with $h$, while its bias increases with $h$. A common and natural approach to find the optimum $h$ value starts by considering the MISE minimization, that is, 
\begin{align}
 \mbox{MISE}(\hat f) &  = \left \langle \int \left ( \hat f(x) - f(x)\right)^2 dx \right \rangle  \nonumber \\ 
& = \int \left ( \langle \hat f(x) \rangle - f(x) \right )^2 dx + \int \mbox{var} \hat f(x) dx \\
& = \int \left [ \left (\mbox{bias} \hat f(x) \right)^2  +  \mbox{var} \hat f(x) \right ]dx\, , \nonumber
\end{align}
where $\langle \;  \; \rangle$ denotes the expected value. There is no simple general solution to perform the above minimization, since the MISE depends on the unknown distribution $f(x)$. 

For simplicity, let both $K$ and $f(x)$ be normal distributions. The kernel can be fixed as such without relevant loss of efficiency, as previously comment. For the second, this is a particular case, but it can be useful as a starting point.  In this context, let $n$ be the number of data points and $\mbox{var}f(x) = \sigma^2$, then  the MISE minimization provides the optimum $h$ as
\begin{equation}
	h_\mscript{opt} = \left(\frac{4}{3}\right)^{1/5} \sigma n^{-1/5}\, . \label{hGaussianRule}
\end{equation}
Therefore, if $f(x)$ can be assumed to be a normal distribution, from the data standard deviation (SD) one can estimate $\sigma$ and use the above formula to find $h_\mscript{opt}$.

For distributions that are close to Gaussian, the rule \eqref{hGaussianRule} should be sufficient. The Silverman rule improves this setting by considering small changes that make it more robust against distributions that are not so close to Gaussian. In particular, the SD-based $h_\mscript{opt}$ formula \eqref{hGaussianRule} tends to generate too large bandwidths for multimodal distributions.  The latter case behaves better with bandwidths based on a more robust dispersion estimate, namely the interquartile range (IQR), which can be defined by the difference between the third and first quartile limits, thus the IQR determines a range that contains about 50\% of the data. For Gaussian data, SD and $\sigma$ are equal, but the IQR is larger than the SD by a factor about 1.34. On the other hand, SD works better than IQR for some distributions, like the binomial one. Following Silverman proposal, the idea is to use an $h_\mscript{opt}$ rule based on the quantity $A$, with
\begin{equation}
	A = \min (\mbox{SD}, \mbox{IQR}/1.34) \, .
\end{equation}
By considering MISE evaluations with this $A$-based rule for several distributions, including some mildly multimodal ones, it was found that on average it is better to replace the factor $(4/3)^{1/5} \sim 1.06$ by $0.9$.  Putting all together, we find the one-dimensional Silverman rule as \citep{Silverman_1998}
\begin{equation}
	h_\mscript{Silverman} = 0.9 \, A \, n^{-1/5} \, .
\end{equation}

The multidimensional case is an extension of the above for $d$ dimensions. It reads
\begin{equation}
	h_\mscript{Silverman} = \frac{9}{10} \left( \frac{ 3}{4} \right)^{\frac{1}{5}} \left (\frac{4}{d+2}\right)^{\frac{1}{d+4}} \, A \, n^{-\frac{1}{d+4}} \, .
\end{equation}

We use the expression above for deriving all the bandwidths.

As a final improvement here implemented, since the probability of finding data with $\rn<0$ or $\rn>1$ is necessarily zero, the smooth PDF found from the KDE method is cropped at $\rn=0$ and $\rn=1$, and then normalized. The complete implementation of such  details can be found in our code \texttt{NAVanalysis}.


\section{Individual fits results}\label{app:IndividualFits}
\setcounter{table}{0} 

Tables~\ref{tab:arctanHalf}, \ref{tab:arctan}, \ref{tab:NFW}, \ref{tab:Burkert}, \ref{tab:MOND} and \ref{tab:MONDdist}  show  the individual fits results from the models Arctan$_{\alpha=1}$, Arctan$_{\alpha=1/2}$, NFW, Burkert, MOND with fixed distance and MOND with Gaussian prior on the distance. All the models consider Gaussian priors on the stellar mass-to-light ratios that are centered on $\YD = 0.5$ and $\YB = 0.6$. 

The tables presented here only contain the first four galaxies of the 153 galaxies  sample, as sorted in alphabetical order. These four galaxies are not supposed to be representative. The full tables are provided in machine readable format as supplementary material.

All the tables display the values of $\chi^2$ and $\chi^2_\mscript{eff}$. The latter was defined in eq.~\eqref{chi2eff}. It is essentially the negative of the logarithm of the posterior. The best fit results are found from the maximization of the posterior, hence by minimizing $\chi^2_\mscript{eff}$. See Sec.~\ref{sec:individual} for further details.

\setlength{\tabcolsep}{10pt}
\begin{table*}
\caption{Arctan$_{1/2}$ individual fits for 153 SPARC galaxies with Gaussian priors on $\YD$ and $\YB$. $r_\mscript{t}$ and $V_\mscript{c}$ are respectively in units of kpc and km/s. The full table is provided in the supplementary material. Galaxies are sorted in alphabetical order.}
\label{tab:arctanHalf}
\begin{center}
\begin{tabular}{l rrrrrrrr}
\hline
\hline\\[-.3cm]
  Galaxy  & 
  \multicolumn{1}{c}{$r_\mscript{t}$} &  
  \multicolumn{1}{c}{$r_\mscript{tn}$} & 
  \multicolumn{1}{c}{$V_\mscript{c}$} & 
  \multicolumn{1}{c}{$\YD$} & 
  \multicolumn{1}{c}{$\YB$} & 
  \multicolumn{1}{c}{$\chi^2$} & 
  \multicolumn{1}{c}{$\chi^2_\mscript{eff}$} & 
  \multicolumn{1}{c}{$N$} \\[.1cm]
\hline
Camb    & 500.00  & 279.33  & 189.13  & 0.36  & ---  & 36.53 & 38.73 & 9\\
D512-2  & 2.61    & 0.68    & 44.38   & 0.50  & ---  & 0.50  & 0.50  & 4\\
D564-8  & 500.00  & 162.87  & 374.50  & 0.48  & ---  & 3.57  & 3.59  & 6\\
D631-7  & 500.00  & 69.54   & 579.26  & 0.30  & ---  & 70.16 & 75.12 & 16\\
$\cdots$ & $\cdots$ & $\cdots$ & $\cdots$ &$\cdots$ &$\cdots$ &$\cdots$ &$\cdots$ & $\cdots$\\ 
\hline 
\hline
\end{tabular}
\end{center}
\end{table*}

\setlength{\tabcolsep}{10pt}
\begin{table*}
\caption{Arctan$_1$ individual fits for 153 SPARC galaxies with Gaussian priors on $\YD$ and $\YB$. $r_\mscript{t}$ and $V_\mscript{c}$ are respectively in units of kpc and km/s. The full table is provided in the supplementary material. Galaxies are sorted in alphabetical order.}
\label{tab:arctan}
\begin{center}
\begin{tabular}{l rrrrrrrr}
\hline
\hline\\[-.3cm]
  Galaxy  & 
  \multicolumn{1}{c}{$r_\mscript{t}$} &  
  \multicolumn{1}{c}{$r_\mscript{tn}$} & 
  \multicolumn{1}{c}{$V_\mscript{c}$} & 
  \multicolumn{1}{c}{$\YD$} & 
  \multicolumn{1}{c}{$\YB$} & 
  \multicolumn{1}{c}{$\chi^2$} & 
  \multicolumn{1}{c}{$\chi^2_\mscript{eff}$} & 
  \multicolumn{1}{c}{$N$} \\[.1cm]
\hline
CamB    & 83.74   & 46.78   & 1000.00 & 0.30  &	---   & 16.84 & 21.74 & 9\\
D512-2  & 0.98    & 0.26    & 41.10   & 0.50  & ---   & 0.23  & 0.23  & 4 \\ 
D564-8  & 1.47    & 0.48    & 33.13   & 0.50  & ---   & 0.24  & 0.24  & 6 \\
D631-7  & 4.82    & 0.67    & 95.96   & 0.39  & ---   & 10.30 & 11.36 & 16 \\
$\cdots$ & $\cdots$ & $\cdots$ & $\cdots$ &$\cdots$ &$\cdots$ &$\cdots$ &$\cdots$ & $\cdots$\\ 
\hline 
\hline
\end{tabular}
\end{center}
\end{table*}

\setlength{\tabcolsep}{10pt}
\begin{table*}
\caption{NFW individual fits for 153 SPARC galaxies with Gaussian priors on $\YD$ and $\YB$. $r_\mscript{s}$ and $\rho_\mscript{s}$ are respectively in units of kpc and M$_\odot$/kpc$^3$. The full table is provided in the supplementary material. Galaxies are sorted in alphabetical order.}
\label{tab:NFW}
\begin{center}
\begin{tabular}{l rrrrrrrr}
\hline
\hline\\[-.3cm]
  Galaxy  & 
  \multicolumn{1}{c}{$r_\mscript{s}$} &  
  \multicolumn{1}{c}{$r_\mscript{sn}$} & 
  \multicolumn{1}{c}{$\rho_\mscript{s}$} & 
  \multicolumn{1}{c}{$\YD$} & 
  \multicolumn{1}{c}{$\YB$} & 
  \multicolumn{1}{c}{$\chi^2$} & 
  \multicolumn{1}{c}{$\chi^2_\mscript{eff}$} & 
  \multicolumn{1}{c}{$N$} \\[.1cm]
\hline
CamB    & 9980.13 & 5575.49 & 2.23  & 0.36  & --- & 36.53 & 38.74 & 9 \\ 
D512-2  & 8.44    & 2.20    & 6.37  & 0.50  & --- & 0.62  & 0.63  & 4 \\ 
D564-8  & 9997.01 & 3256.36 & 2.82  & 0.48  & --- & 3.58  & 3.60  & 6 \\ 
D631-7  & 9997.22 & 1390.43 & 3.20  & 0.30  & --- & 70.23 & 75.19 & 16 \\ 
$\cdots$ & $\cdots$ & $\cdots$ & $\cdots$ &$\cdots$ &$\cdots$ &$\cdots$ &$\cdots$ & $\cdots$\\ 
\hline 
\hline
\end{tabular}
\end{center}
\end{table*}

\setlength{\tabcolsep}{10pt}
\begin{table*}
\caption{Burkert individual fits for 153 SPARC galaxies with Gaussian priors on $\YD$ and $\YB$. $r_\mscript{c}$ and $\rho_\mscript{c}$ are respectively in units of kpc and M$_\odot$/kpc$^3$. The full table is provided in the supplementary material. Galaxies are sorted in alphabetical order.}
\label{tab:Burkert}
\begin{center}
\begin{tabular}{l rrrrrrrr}
\hline
\hline\\[-.3cm]
  Galaxy  & 
  \multicolumn{1}{c}{$r_\mscript{c}$} &  
  \multicolumn{1}{c}{$r_\mscript{cn}$} & 
  \multicolumn{1}{c}{$\rho_\mscript{c}$} & 
  \multicolumn{1}{c}{$\YD$} & 
  \multicolumn{1}{c}{$\YB$} & 
  \multicolumn{1}{c}{$\chi^2$} & 
  \multicolumn{1}{c}{$\chi^2_\mscript{eff}$} & 
  \multicolumn{1}{c}{$N$} \\[.1cm]
\hline
CamB    & 1000.00 & 558.66  & 6.51  & 0.30  & --- & 16.85 & 21.75 & 9 \\
D512-2  & 1.50  & 0.39      & 7.65  & 0.50  & --- & 0.12  & 0.12  & 4 \\
D564-8  & 2.15  & 0.70      & 7.12  & 0.50  & --- & 0.27  & 0.27  & 6 \\
D631-7  & 7.00  & 0.97      & 7.02  & 0.39  & --- & 11.27 & 12.48 & 16 \\
$\cdots$ & $\cdots$ & $\cdots$ & $\cdots$ &$\cdots$ &$\cdots$ &$\cdots$ &$\cdots$ & $\cdots$\\ 
\hline 
\hline
\end{tabular}
\end{center}
\end{table*}

\setlength{\tabcolsep}{10pt}
\begin{table*}
\caption{MOND individual fits for 153 SPARC galaxies with Gaussian priors on $\YD$ and $\YB$. Galaxy distance is fixed and it is assumed $a_0 = 1.2 \times 10^{-13}$km/s$^2$. The full table is provided in the supplementary material. Galaxies are sorted in alphabetical order.}
\label{tab:MOND}
\begin{center}
\begin{tabular}{l rrrrrrrr}
\hline
\hline\\[-.3cm]
  Galaxy  & 
  \multicolumn{1}{c}{$\YD$} & 
  \multicolumn{1}{c}{$\YB$} & 
  \multicolumn{1}{c}{$\chi^2$} & 
  \multicolumn{1}{c}{$\chi^2_\mscript{eff}$} & 
  \multicolumn{1}{c}{$N$} \\[.1cm]
\hline
CamB    & 0.06 & --- & 405.10   & 492.00  & 9 \\
D512-2  & 0.35 & --- & 3.00     & 5.36    & 4 \\ 
D564-8  & 0.20 & --- & 53.34    & 69.49   & 6 \\ 
D631-7  & 0.11 & --- & 272.27   & 313.87  & 16 \\ 
$\cdots$ & $\cdots$ & $\cdots$ & $\cdots$ &$\cdots$ &$\cdots$ \\ 
\hline 
\hline
\end{tabular}
\end{center}
\end{table*}

\setlength{\tabcolsep}{10pt}
\begin{table*}
\caption{MOND individual fits for 153 SPARC galaxies with Gaussian priors on $\YD$, $\YB$ and the galaxy distance ($D$). $\delta D$ is a dimensionless factor that multiplies the most probable galaxy distance ($D_0$, hence $D = D_0 \, \delta D$). It is assumed $a_0 = 1.2 \times 10^{-13}$km/s$^2$. The full table is provided in the supplementary material. Galaxies are sorted in alphabetical order.}
\label{tab:MONDdist}
\begin{center}
\begin{tabular}{l rrrrrrrr}
\hline
\hline\\[-.3cm]
  Galaxy  & 
  \multicolumn{1}{c}{$\YD$} & 
  \multicolumn{1}{c}{$\YB$} & 
  \multicolumn{1}{c}{$\delta D$} & 
  \multicolumn{1}{c}{$\chi^2$} & 
  \multicolumn{1}{c}{$\chi^2_\mscript{eff}$} & 
  \multicolumn{1}{c}{$N$} \\[.1cm]
\hline
CamB    & 0.30  & --- & 0.28  & 37.68   & 130.34  & 9 \\ 
D512-2  & 0.48  & --- & 0.76  & 0.37    & 1.07    & 4 \\ 
D564-8  & 0.23  & --- & 0.90  & 36.19   & 57.84   & 6 \\ 
D631-7  & 0.16  & --- & 0.86  & 184.95  & 245.53  & 16 \\
$\cdots$ & $\cdots$ & $\cdots$ & $\cdots$ &$\cdots$ &$\cdots$ &$\cdots$\\ 
\hline 
\hline
\end{tabular}
\end{center}
\end{table*}

\section{NAV analysis and the $M_{200}-c$ plane for NFW} \label{app: M200-c}
\setcounter{figure}{0} 

Here we consider an application of the NAV analysis to the halo mass-contentration relation ($M_{200}-c$). It is well known that the NFW profile has a tendency of finding best-fits with unrealistically high values for $r_\mscript{s}$, when considered with flat priors on $\rho_s$ and $r_s$. The NFW fits done also show this behaviour (Fig.~\ref{fig:plotNFWburkert}). In the $M_{200}-c$ plane, this behaviour leads to virial masses too high and concentrations too low, outside the expectations from the cosmological simulations \citep[e.g.,][]{deBlok:2002tg, Rodrigues:2014xka, Li:2020iib}.

In Sec.~\ref{sec:NAVNFW}, NFW constraints were found for the parameter $r_\mscript{sn}$. These are also constraints on $M_{200}$ and $c$, since $r_\mscript{sn}$ can be written as a function of the latter. Indeed, following \citet{0521857937},
\begin{align}
 & c = r_\mscript{s}/r_{200} \, , \\
 & M_{200} = M(r_{200}) = 4 \pi \Omega_m \rho_\mscript{s} r^3_\mscript{s} \left( \ln(1+c) - \frac{c}{1+c}\right) \, ,  \nonumber \\
& \rho_\mscript{s} = 200 \Omega_\mscript{m} \rho_\mscript{c} c (1+c)^2 \, , \nonumber
\end{align}
where $\rho_\mscript{c}$ is the cosmological critical density.

Hence $r_\mscript{sn}$ can be written as a function of $M_{200}$ and $c$ as 
\begin{equation}
	r_\mscript{sn} = \left( \frac{M_{200}}{4 \pi \rho_\mscript{s} \Omega_\mscript{m} \left( \ln(1+c) - \frac{c}{1+c} \right )}   \right )^{1/3} \frac{1}{r_\mscript{max}} \, .
\end{equation}

\begin{figure}
	\begin{tikzpicture}
  		\node (img1)  {\includegraphics[width=0.45\textwidth]{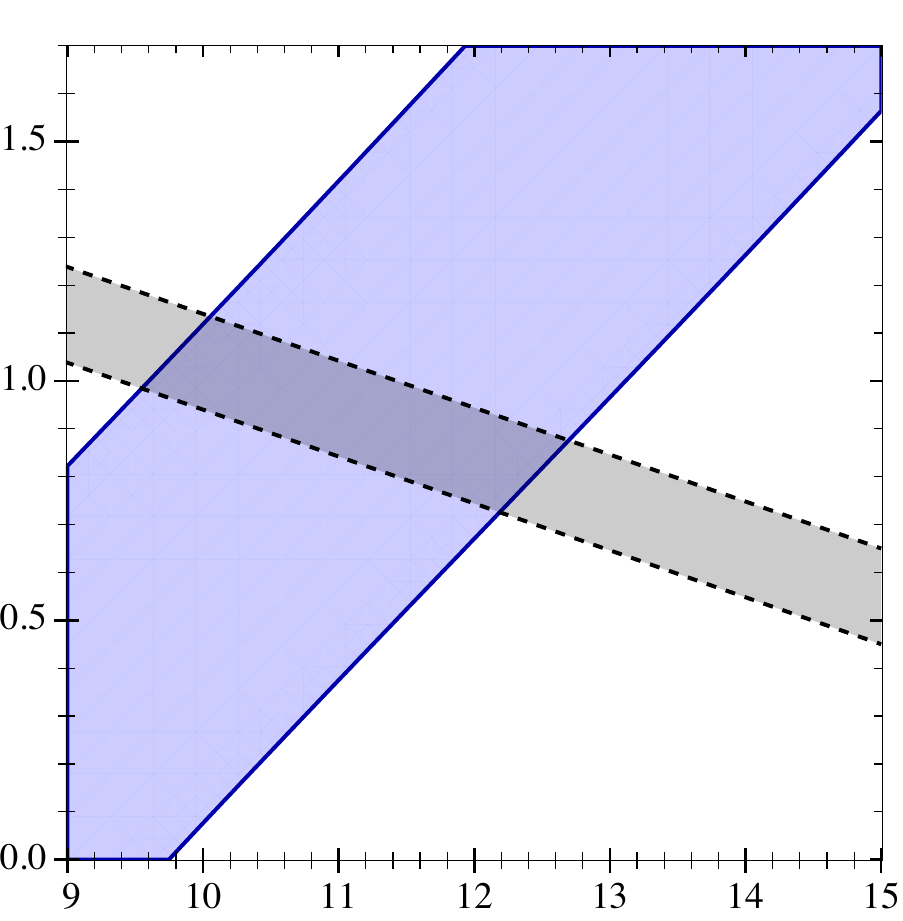}};
		\node[below=of img1, node distance=0cm, yshift=1.1cm, xshift=0.3cm, font=\color{black}] {\large $\log M_{200} (M_\odot)$};
		\node[left=of img1, node distance=0cm, rotate=90, yshift=-0.8cm, xshift=0.6cm] {\large $\log c$};
	\end{tikzpicture}
	\caption{The gray-dashed region shows an approximation for the $M_{200}-c$ correlation, at 1$\sigma$, as derived from cosmological N-body simulations by \citet{Maccio:2008pcd}. The plotted ranges of $M_{200}$ and $c$ are the same of the previous reference. The blue region, which is delimited by a solid line, shows the region compatible with the $1\sigma$ NFW constraint on $r_\mscript{sn}$ \eqref{rsnHDRLimits}.}
	\label{fig:plotM200-c200} 
\end{figure}

For a given $r_\mscript{max}$ value, the constraints from eq.~\eqref{rsnHDRLimits} will yield a region in the $M_{200}-c$ plane. Using the median of $r_\mscript{max}$ from the 153 SPARC galaxies, which is 12.37 kpc, one finds the $1\sigma$ constraint in the $M_{200}-c$ plane as shown blue in Fig.~\ref{fig:plotM200-c200}. Higher $r_\mscript{max}$ values slightly shifts this region towards higher $M_{200}$ values, while lower $r_\mscript{max}$ shifts the region in the opposite direction.  This blue region alone is independent from any expectations from cosmological simulations, it was derived considering only the NFW halo radial dependence when contrasted to the observational data. Most of the individual NFW fits (with flat priors on $r_\mscript{s}$ and $\rho_\mscript{s}$) are expected and indeed appear in this region, since this region is only a reparametrization of the 1$\sigma$ $r_\mscript{sn}$ constraint \citep[see also][]{Li:2020iib}.

The bottom-right region of Fig.~\ref{fig:plotM200-c200} is not covered by the blue region, but  the former region is  covered at the 2$\sigma$ level (since at this level there is no upper limit for $r_\mscript{sn}$). Any constraints imposed such that this bottom-right region is not populated by galaxy fits will hinder NFW fits that yield large $r_\mscript{s}$ or $r_\mscript{sn}$. This is the case of imposing constraints from cosmological simulations (Fig.~\ref{fig:plotM200-c200}). The effect of such constraints can be seen in the NAV plane: they will not allow the NFW contribution to the rotation curve to become close to its lower limit (Fig.~\ref{fig:plotNFWburkert}), hence the NFW halo with these constraints will have an additional difficulty on covering observational data, which in general will lead to higher $\chi^2$ values. 

As it can be inferred from Fig.~\ref{fig:plotNFWburkert}, the extra difficulty in covering the observational data, due to cosmological constraints in the $M_{200}-c$ plane,  does not necessarily imply high $\chi^2$ penalties. This happens since, for NFW curves that are close to the limiting case $\delta V^2_\mscript{NFW} = r_\mscript{n}$, $r_\mscript{sn}$ may change by orders of magnitude without a relevant change in $\delta V^2_\mscript{NFW}$.

\bibliographystyle{elsarticle-harv}

\end{document}